\documentclass[aoas]{imsart}

\RequirePackage{amsthm,amsmath,amsfonts,amssymb}
\RequirePackage[colorlinks,citecolor=blue,urlcolor=blue]{hyperref}
\RequirePackage{graphicx}
\RequirePackage{ulem}

\startlocaldefs

\newtheorem{theorem}{Theorem}[section]

\theoremstyle{remark}

\usepackage{color}
\usepackage{indentfirst}
\usepackage{amsmath}
\usepackage{bm}
\usepackage{graphicx}
\usepackage{float}
\usepackage{algorithm,algcompatible}
\usepackage{caption}
\usepackage{makecell}
\usepackage{ulem,soul}
\algnewcommand\INPUT{\item[\textbf{Input:}]}%
\algnewcommand\OUTPUT{\item[\textbf{Output:}]}%

\newcommand\numberthis{\addtocounter{equation}{1}\tag{\theequation}}
\DeclareMathOperator*{\argmin}{arg\,min}

\algdef{SE}[SUBALG]{Indent}{EndIndent}{}{\algorithmicend\ }%
\algtext*{Indent}
\algtext*{EndIndent}
\endlocaldefs
\renewcommand\thesection{\Alph{section}}

\begin{document}

\begin{frontmatter}
\title{Matrix Completion Methods for the Total Electron Content Video Reconstruction}
\runtitle{Matrix Completion for TEC Reconstruction}

\begin{aug}
\author[A]{\fnms{Hu} \snm{Sun}},
\author[A]{\fnms{Zhijun} \snm{Hua} },
\author[B]{\fnms{Jiaen} \snm{Ren} },
\author[B]{\fnms{Shasha} \snm{Zou} },
\author[A]{\fnms{Yuekai}
\snm{Sun}},
\and 
\author[A,C]{\fnms{Yang} \snm{Chen}\thanks{Corresponding author.}\ead[label=e1]{ychenang@umich.edu} }
\address[A]{Department of Statistics,
University of Michigan, Ann Arbor
\printead{e1}}

\address[B]{Department of Climate and Space Sciences and Engineering,
University of Michigan, Ann Arbor}

\address[C]{Michigan Institute for Data Science (MIDAS), University of Michigan, Ann Arbor}
\end{aug}

\begin{abstract}%

The total electron content (TEC) maps can be used to estimate the signal delay of GPS due to the ionospheric electron content between a receiver and satellite. This delay can result in GPS positioning error. Thus it is important to monitor the TEC maps. The observed TEC maps have big patches of missingness in the ocean and scattered small areas of missingness on the land. In this paper, we propose several extensions of existing matrix completion algorithms to achieve TEC map reconstruction, accounting for spatial smoothness and temporal consistency while preserving important structures of the TEC maps. We call the proposed method Video Imputation with SoftImpute, Temporal smoothing and Auxiliary data (VISTA). Numerical simulations that mimic patterns of real data are given. We show that our proposed method achieves better reconstructed TEC maps as compared to existing methods in literature. Our proposed computational algorithm is general and can be readily applied for other problems besides TEC map reconstruction. 
\end{abstract}


\end{frontmatter}

\section{Introduction}

\subsection{Introduction to TEC Maps}

``Space Weather” refers to the variable conditions on the Sun and in the space environment that can adversely influence the performance and reliability of space-borne and ground-based technological systems \cite{NRC1997}. In recent years, there has been a growing awareness of space weather impacts on critical infrastructure in the civilian, commercial, and military sectors. Understanding the underlying physical processes of space weather and improving the forecasting is a major objective of the space science community. 

The ionosphere, a layer in the upper atmosphere that extends from ~70 km to ~1000 km above the Earth’s surface, contains roughly equal number of electrons and ions, which are mainly produced by the ionization of the neutral atmosphere by solar UV radiation and impact ionization by precipitating energetic particles from space. Ionospheric state and variability depend on solar activity, near-Earth space environment condition, time of day and day of year, as well as geographic location. Eruptive space weather events, such as coronal mass ejections (CMEs), have the largest impact on ionospheric state and its variability \cite{Mendillo2006, Prolss2008, Zou2013, Zou2014}. Ionospheric disturbances are highlighted as one out of the five major space weather threats in the National Space Weather Strategy and Action Plan \cite{NSWSAP2019}.

The Global Navigation Satellite Systems (GNSS) systems are initially designed for Positioning, Navigation and Timing (PNT) services, but have also been widely used in the space science community for remotely sensing the ionosphere total electron content (TEC). TEC refers to the integrated electron density between the receivers and the GNSS satellites and can be calculated using the different delays of two or more transmitted frequencies from multi-frequency GNSS satellites. In order to achieve PNT accuracy, single frequency GNSS receivers on the ground need the ionosphere TEC information to remove the ionosphere impact. For example, Federal Aviation Agency (FAA) developed the Aide Area Augmentation System (WAAS) to estimate and augment the ionosphere delay in order to improve the PNT service accuracy for aviation. Therefore, specification and forecasting the ionosphere plasma content, i.e., the TEC map, and its variability are of critical importance to our modern technological society.

\begin{figure}[h]
    \centering
    \includegraphics[width=0.98\textwidth]{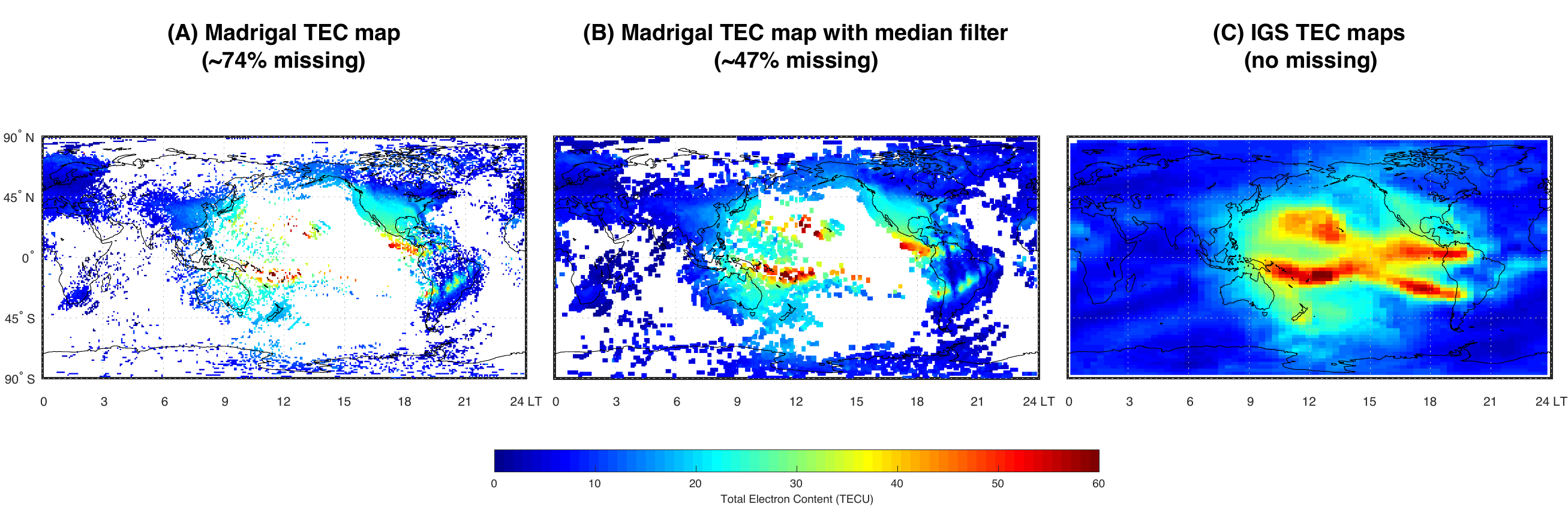}
    \caption{TEC map from the Madrigal Database (A) without median filter on the left, (B) with a ${3^\circ \times 3^\circ}$ median filter on the right and (C) TEC map from the International GNSS Service (IGS).}
    \label{fig:madrigal}
\end{figure}

The Madrigal Database~ \cite{Rideout2006,Vierinen2016} provides global maps of vertical TEC measurements calculated from dual-frequency GNSS data collected by world-wide distributed receivers (over 5000 of them). The GNSS system used here include both the Global Positioning System (GPS) and the Global Navigation Satellite System (GLONASS). The Madrigal TEC maps are provided with a spatial resolution of $1^{\circ}$ latitude by $1^{\circ}$ longitude and a temporal resolution of 5 minutes. Mainly due to the absence of GNSS receivers over the oceans, the TEC measurement is missing at around 75\% of the globe (see the left panel on Figure \ref{fig:madrigal}). In practice, a ${3^\circ}$-by-${3^\circ}$ median filter is usually applied to reduce the percentage of missing values down to about 50\%. In this paper, we adopt the median filtered data as the observations and seek for TEC map reconstruction based on these maps with about 50\% missing values (see the right panel on Figure~\ref{fig:madrigal}).

Complete global ionospheric maps (GIMs) of vertical TEC values provided by the International GNSS Service (IGS) are produced by combining TEC maps calculated by several IGS Ionosphere Associated Analysis Centers (IAACs), which use different techniques but mostly expansion in terms of spherical harmonics (SH) to model the global VTEC maps \cite{Schaer1999,Hernandez-Pajares2009,Roma-Dollase2018}. The IAACs use TEC data collected by around 200-500 IGS GNSS receivers. The commonly used technique models the global TEC map by a expansion consisting of SH functions, whose coefficients are obtained by fitting the measured TEC data based on the least squared algorithm \cite{Schaer1995}. Additional constrains can be applied to the fitting process to improve the resulting model, e.g. removing negative TEC values by adding an inequality constraint \cite{Zhang2013}. The TEC GIMs provided by IGS have a spatial resolution of $2.5^{\circ}$ latitude by $5.0^{\circ}$ longitude and a highest temporal resolution of 15 minutes. As shown in the right panel of Figure \ref{fig:madrigal}, the IGS TEC maps can well approximate the TEC distribution at a global scale, however, meso-scale TEC structures are generally smoothed out in these fitted models, while they are important for ionospheric scientific research and for augmenting the ionospheric impact on the PNT service (e.g.  \cite{conker2003modeling,yang2020global}). 
For example, in Figure 1a and 1b, a clear longitudinally extended low TEC channel at about 19 local time (LT), i.e., equatorial plasma bubble, can be see clearly, while it is smoothed out in Figure 1c. Equatorial plasma bubble is one of the most important ionospheric density and TEC features that can severely degrade the GPS signals or even lead to loss of lock \cite{aa2019,basu2002}. 

\subsection{TEC Map Reconstruction Problem}
\label{subsec:tec_reconstruct_problem_statement}

The objective of this work is to improve upon existing TEC map reconstruction algorithms, namely the Spherical Harmonics method, to obtain maps that comply with the observed values as much as possible while preserving local and global features in the map. 

In this study, the TEC maps are converted into magnetic local time (MLT) coordinates \cite{Shepherd2014}, where the noon (12 MLT) is always fixed at the center of each map while the locations of the continents and oceans are constantly shifting over time with the Earth's rotation. Recall that TEC values over the oceans are largely missing due to the lack of GNSS receivers. Therefore, the pattern of missing values is structured: the TEC matrices have big patches of missingness which move across time.  

Mathematically, the TEC maps over an extended time period (e.g. one day) can be represented by $m\times n$ matrices $\{X_t, t=1, 2,\ldots, T\}$, each of which has missing values; and the locations of the missingness vary across different time points. For any arbitrary matrix $X$, let $\Omega$ denote the observed entries in $X$; i.e. $\Omega = \{ (i, j): X_{ij} \text{ is observed}\}$. Following the notation in \cite{candes2010power}, the projection $\mathrm{P}_{\Omega}(X)$ is an $m\times n$ matrix keeping all observed entries of $X$ and replacing all missing entries with $0$. And $P_\Omega ^\perp $ projects onto the complement of the set $\Omega$. In the next section, we review a fraction of the literature on different formulations of the matrix completion problem. Note that we only discuss the ones that are directly relevant to our work and do not give a full literature review on the subject, see \cite{li2019survey,jain2013low,hastie2015matrix} and references therein for more related works on low-rank matrix completion.

\subsection{Literature Review on Matrix Completion}
\label{subsec:litreview}

We use the notations defined in Section~\ref{subsec:tec_reconstruct_problem_statement}. The early matrix completion method by \cite{mazumder2010spectral} aims at solving the following convex-optimization problem for each $X$:
\begin{equation}\label{Nuclear-Norm}
    \min_{M} \left\{ H(M) := \frac12 \|\mathrm{P}_{\Omega}(X - M)\|^{2}_{F} + \lambda \|M\|_{*}\right\},
\end{equation}
where $\|\cdot\|_{*}$ is the nuclear norm, i.e. sum of all singular values. It is a well-known result that when $X$ is fully-observed, the solution to \eqref{Nuclear-Norm} is $M=U_r\bm{S}_{\lambda}(D_r)V_r^T$, where $r=\min(m,n)$; $U_r,D_r,V_r$ are from the rank-$r$ Singular Value Decomposition (SVD) of $X$, i.e., $X=U_rD_rV_r'$, $D_r = \mathbf{diag}\left[(\sigma_1, \sigma_2, \dots, \sigma_r)\right]$; and $\bm{S}_{\lambda}(D_r) = \mathbf{diag}\left[(\sigma_1-\lambda)_+,(\sigma_2-\lambda)_+,\dots,(\sigma_r-\lambda)_+\right]$ is the soft-thresholding operator. And when $X$ contains missing values, the problem in \eqref{Nuclear-Norm} can be solved by an iterative algorithm. In each step of the algorithm, the missing entries in the matrix is first imputed by the current imputation matrix $\tilde{M}$ to get an ``imputed`` $\tilde{X}$ and then the rank-$r$ SVD is done on $\tilde{X}$ to update $\tilde{M}$ until convergence.

In \cite{hastie2015matrix}, inspired by the maximum-margin matrix factorization (MMMF) formulation in \cite{srebro2005maximum}, a different formulation called the softImpute-ALS, which our method is based on, is proposed. It seeks to find factor matrices $A$ and $B$ such that $AB^T$ approximates the matrix $X$. Iteratively, the following optimizaiton problem is solved:
\begin{equation}\label{softImpute-ALS}
    \min_{A, B} \left\{ F(A, B|\tilde{A}, \tilde{B}) := \frac12 \|\widehat{X} - AB^T\|^{2}_{F} + \frac{\lambda_1}{2} (\|A\|^{2}_{F} + \|B\|^{2}_{F})\right\},
\end{equation}
where $\|\cdot\|_F$ is the Frobenius norm, defined as the square root of the sum of the squares of its elements, $\widehat{X}$ is a ``filled-in" $m\times n$ matrix, with $\widehat{X} = \mathrm{P}_{\Omega}(X) + \mathrm{P}_{\Omega^{\perp}} (\tilde{A}\tilde{B}^T)$, and $\tilde{A}, \tilde{B}$ are the two factor matrices in the previous iterative step. The algorithm works via doing alternating ridge regression to update $A$ and $B$, and eventually, the solution is $\widehat{A}=U_r\bm{S}_{\lambda}(D_r)^{\frac12}$ and $\widehat{B}=V_r\bm{S}_{\lambda}(D_r)^{\frac12}$, which agrees with the algorithm of \eqref{Nuclear-Norm}. 

The nuclear-norm penalty term in \eqref{Nuclear-Norm} and the L2-penalty in \eqref{softImpute-ALS} share a common modeling intention: put a soft constraint on the rank of the imputation matrix, so that the optimization problem is made (bi-)convex. Lemma 1 in \cite{srebro2005maximum} also shows that $\|X\|_{*} = \min_{X=AB^{T}} 
\frac12 \left(\|A\|^{2}_{F} + \|B\|^{2}_{F}\right)$, so the penalty term in \eqref{softImpute-ALS} is a further relaxation of the nuclear-norm penalty. The  $A_{m\times r}$ factor and the $B_{n\times r}$ factor have direct interpretations: the $r$-dimensional latent feature for every row and column of the matrix. Each entry in the final imputation is just the inner product of the latent feature of the corresponding row and column (latitude and longitude in the TEC map). We adopt the factorization setup and the L2-penalty in \eqref{softImpute-ALS} because of its interpretability, direct control of $r$ and its elegant least-square solution. The default of our algorithm in the following text, however, fixes $r=\min(m,n)$. But one can impose a low-rank structure in the first place if prior knowledge is present. See Section 8 of \cite{mazumder2010spectral} for further comparison of the two types of penalty terms.

\begin{figure}[h]
    \centering
    \includegraphics[width=0.98\textwidth]{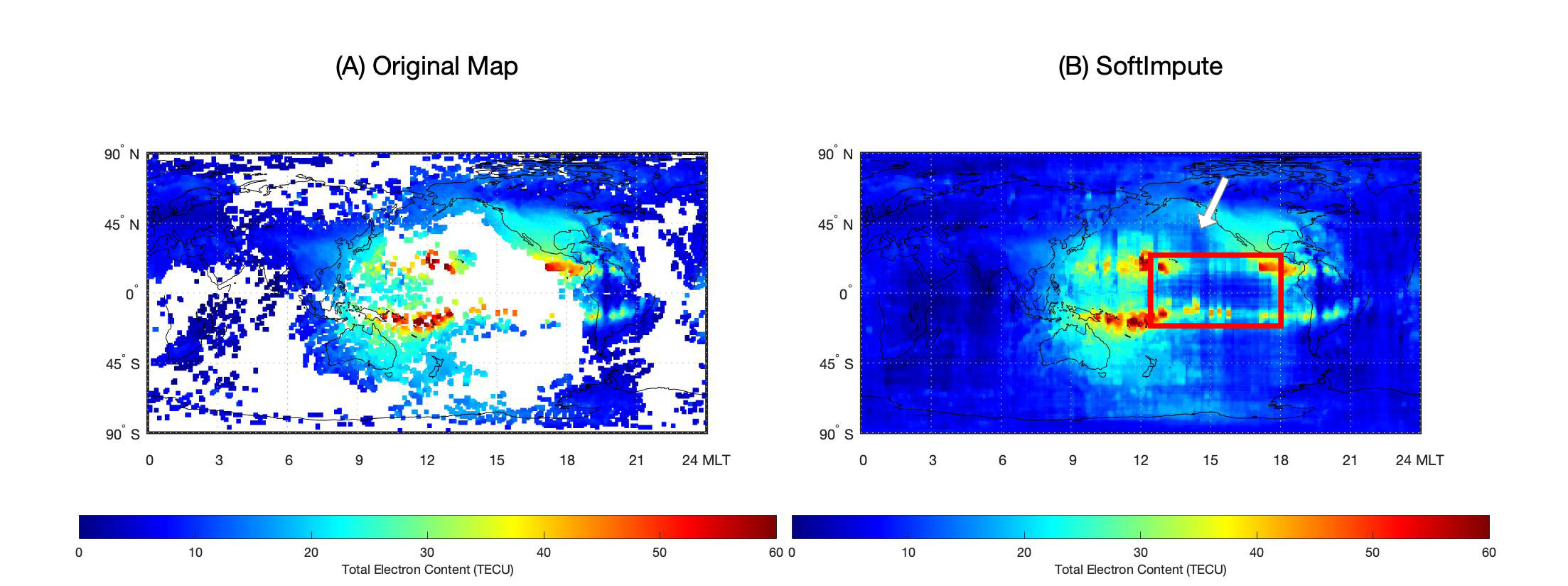}
    \caption{TEC maps: observed (left) and fitted by the SoftImpute approach (right).}
    \label{fig:softimpute_ex}
\end{figure}

In our TEC map reconstruction context, one can naively apply the softImpute-ALS to each TEC matrix $X_t$ separately by iteratively solving \eqref{softImpute-ALS}. In Figure~\ref{fig:softimpute_ex}, we show the observed TEC map on the left and the reconstructed TEC map with the softImpute-ALS method on the right. This figure reveals two problems of applying the original softImpute-ALS algorithm directly to the TEC map reconstruction problem.  
\begin{enumerate}
    \item The imputed TEC map, namely $\widehat{A}\widehat{B}^{T}$, exhibits a \textit{global} {unsmooth} structure which is not ideal.{The maximum possible rank is $181$ for the map but the imputed one generally has rank $70\sim90$.}  {T}here are many rows (or columns) imputed with equal, near-zero values, i.e., the blue bands in the right panel of Figure~\ref{fig:softimpute_ex} at about 15 MLT and $20^{\circ}$ N (pointed at by the white arrow). Such near zero TEC values embedded in the dayside high TEC region (often called equatorial ionization anomaly) is physically unreasonable. A more reasonable imputation should keep the spatial continuity of TEC maps, especially in the dayside, high TEC value sub-regions. {One can lower the L2-penalty in softImpute-ALS to improve the smoothness, thus increasing the rank, of the imputed map; but this results in over-fitting the observed region. We need the extra smoothness penalty to improve the missing region smoothness while not over-fitting the observed regions.}
    \item Large patches of unobserved values in the TEC map are poorly imputed by values near zero (see the red highlighted part in Figure \ref{fig:softimpute_ex}(B).), even though the patches are clearly part of a sub-region with high TEC values. 
    
\end{enumerate}

The SoftImpute and a lot of other matrix completion methods, as we briefly mention a few below (especially work by \cite{Bell2007TheBS}, \cite{koren2008factorization} and \cite{koren2009bellkor}), are designed to deal with the imputation of matrices with scattered, random, non-patch missingness, such as building a recommendation system for \texttt{Netflix Prize} by \cite{bennett2007netflix}. However, our data, TEC maps, are scientific images and have non-random, auto-correlated, and patch missingness. Thus, these existing methods do not perform as desired in the TEC map completion problem.

Other related work include \cite{mao2004modeling} and \cite{lee2000algorithms}, which focused on non-negative matrix completion. \cite{chen2005nonnegative} proposed nonnegative matrix factorization with temporal smoothness, however, their method could not deal with correlated spatial constraints. \cite{barnes2009patchmatch} proposed patch matching algorithm for image completion for nonparametric texture construction. However, the performance was not satisfactory when the original image lacked adequate data to complete the missing regions. \cite{huang2014image} extracted mid-level constraints  and used them to guide the filling of missing regions. Yet the corrupted region must be small and relevant to visual data to have a good completion result. Our proposed method in this paper, on the contrary, is capable of imputing a time-series of matrices with a large amount of missing values and guarantee both spatial smoothness and temporal consistency, which turns out to be very helpful for reconstructing scientific images, i.e., the TEC maps/videos.

\section{Methodology}\label{method-section}

In this section, we first summarize the two extensions we add on top of the softImpute-ALS method for our TEC map reconstruction task. We call the proposed method Video Imputation with SoftImpute, Temporal smoothing and Auxiliary data (VISTA). In section \ref{algorithm-section}, we present the details of the algorithm. Convergence properties of the algorithm are discussed in section \ref{theory-section}.

Following the notations we defined in Section \ref{subsec:tec_reconstruct_problem_statement}, we consider a set of $m\times n$ matrices $\{X_1,X_2,\dots,X_T\}$, each of which has missing values to be imputed. For each matrix $X_t$, $1\leq t\leq T$, the observed entries are indexed by the set $\Omega_t = \{(i,j): X_{t}(i,j) \mbox{ is observed}\}$ and the missing entries are indexed by the complement $\Omega_t^{\perp}$. 

\subsection{Extensions of the softImpute Method}

To improve the softImpute-ALS matrix completion method with the aim of addressing the two issues described in Section \ref{subsec:litreview}, we propose a more general framework for matrix (video) completion, based on the softImpute-ALS algorithm and including it as a special case. In summary, our method is solving the following optimization problem:

\begin{align*}
    \min_{A_{1:T},B_{1:T}} \bigg\{ F(A_{1:T},B_{1:T}) &\triangleq \frac12\sum_{t=1}^T\|P_{\Omega_t}(X_t - A_{t}B_{t}^T)\|_F^2 + \frac{\lambda_1}{2}\sum_{t=1}^T(\|A_t\|_F^2 + \|B_t\|_F^2) \\
    & \quad + \frac{\lambda_2}{2}\sum_{t=2}^T\|A_{t}B_{t}^T - A_{t-1}B_{t-1}^T\|_F^2 + \frac{\lambda_3}{2} \sum_{t=1}^T \|Y_t - A_{t}B_{t}^T\|_F^2\bigg\}, \numberthis{} \label{softALS-TSSH}
\end{align*}
where $Y_1,Y_2,\dots,Y_T$ are $m\times n$ auxiliary data with no missing values. Auxiliary data are obtained by applying some smoothing function over $X_t$, which typically results in overly smoothing the data thus not complying with the observations as desired. In our TEC map reconstruction, we smooth each TEC map $X_t$ with Spherical Harmonics (SH) with relatively low orders of complexity, and the smoothed SH data are used as auxiliary data. In fact, one can use the imputed maps out of an arbitrary imputation algorithm as the auxiliary data. We pick Spherical Harmonics for its spatial smoothness. Backgrounds about Spherical Harmonics will be briefly introduced in section \ref{subsec:SH}, and one can just think the auxiliary data as a series of fully imputed maps for now. Note that our algorithm allows having this auxiliary data if such data exists and is justified within the scientific field. However, if such data is not available, we can simply set $\lambda_3=0$. 

In \eqref{softALS-TSSH}, we add two additional terms on top of \eqref{softImpute-ALS}, each serves as a solution to the two problems of softImpute-ALS. The term with $\lambda_2$ is introducing temporal-smoothing (TS) to the imputation. This enables information sharing across neighboring time points. The term with $\lambda_3$ is making the imputation to learn from both the original observed data $\{X_t\}$ and the auxiliary data $\{Y_t\}$. Dropping both terms reduces the optimization problem to that in \eqref{softImpute-ALS}, which is solved via the original softImpute-ALS algorithm.

The reason for including the temporal-smoothing is that sub-regions with relatively high values tend to remain stable in adjacent frames, and penalizing the difference of imputed matrices between adjacent frames can eliminate undesirable low-rank structure of the imputed sub-regions. The reason for learning from the auxiliary data is that the auxiliary data, such as the reconstructed map fitted with Spherical Harmonics, has reasonable ``observations" in the large missing patches, thus providing additional information for imputing the large sub-regions that nearly $100\%$ missing. This is a special pattern of missingness that we face in the TEC map/video reconstruction task. 


\subsection{Description of Algorithms}\label{algorithm-section}
In the previous section, we set up our matrix completion problem as an optimization problem \eqref{softALS-TSSH}. 
Although it is possible to solve \eqref{softALS-TSSH} directly with off-the-shelf solvers, such solvers do not scale readily to large-scale problems. Following the approach \cite{hastie2015matrix}, we develop a majorization-minimization (MM) approach to solve \eqref{softALS-TSSH} at scale. The main difference from \cite{hastie2015matrix} is that we have a sequence of matrices that we wish to complete jointly, so their algorithm must be modified, and its justification re-established. And we give the details of the modifications in this section. 

At a high-level, the method is a optimized alternating least square (ALS) approach. 
The ALS method is commonly seen in matrix factorization \cite{paatero1994positive,kim2008nonnegative,kim2008toward} and matrix completion \cite{koren2009matrix,gimenez2019matrix}. Since we have more than two matrices, we update the factors $A_1, A_2, \dots, A_T, B_1, B_2, \dots, B_T$ cyclically: $A_1 \rightarrow A_2 \rightarrow \dots \rightarrow A_T \rightarrow B_1 \rightarrow B_2 \rightarrow \dots \rightarrow B_T \rightarrow A_1 \rightarrow A_2 \rightarrow \dots$.
This is an instance of cyclic block coordinate descent \cite{xu2013block}. In each step we update one factor, keeping the other $2T-1$ factors at their current values. 

Suppose in the $k$-th round, we wish to update $A_t$. The current values for the other factors are: $A_{1}^{(k+1)}, A_{2}^{(k+1)}, \dots, A_{t-1}^{(k+1)}, A_{t}^{(k)}, \dots  A_{T}^{(k)}$ and $B_{1}^{(k)}, B_{2}^{(k)}, \dots, B_{T}^{(k)}$. 
Keeping every matrix other than $A_t$ fixed at their current values, the convex optimization problem in \eqref{softALS-TSSH} is reduced to the following optimization problem:
\begin{align*}
    &\min_{A_{t}} 
    \bigg\{ Q(A_{t}|A_{1:t-1}^{(k+1)},A_{t+1:T}^{(k)},B_{1:T}^{(k)})\\
    &\triangleq \frac12\|P_{\Omega_t}(X_t - A_{t}(B_{t}^{(k)})^T)\|_F^2 + \frac{\lambda_1}{2}\|A_t\|_F^2 + \frac{\lambda_3}{2} \|Y_t - A_{t}(B_{t}^{(k)})^T\|_F^2 \\
    &\quad + \frac{\lambda_2}{2}\mathbf{I}_{\{t> 1\}}\|A_{t}(B_{t}^{(k)})^T - A_{t-1}^{(k+1)}(B_{t-1}^{(k)})^T\|_F^2 \\
    &\quad + \frac{\lambda_2}{2}\mathbf{I}_{\{t< T\}}\|A_{t+1}^{(k)}(B_{t+1}^{(k)})^T - A_{t}(B_{t}^{(k)})^T\|_F^2 \bigg\}, \numberthis{} \label{surrogate-A}
\end{align*}
where $\mathbf{I}_{\left\{\cdot\right\}}$ is an indicator function which is equal to 1 if the condition in the subscript holds and zero otherwise. Similarly for each $B_t$, by keeping all factors other than $B_t$ constant, the optimization problem in \eqref{softALS-TSSH} is reduced to the following optimization problem:
\begin{align*}
    &\min_{B_{t}} \bigg\{Q(B_{t}|A_{1:T}^{(k+1)},B_{1:t-1}^{(k+1)},B_{t+1:T}^{(k)}) \\
    &\triangleq \frac12\|P_{\Omega_t}(X_t - A_{t}^{(k+1)}B_{t}^T)\|_F^2 + \frac{\lambda_1}{2}\|B_t\|_F^2 + \frac{\lambda_3}{2} \|Y_t - A_{t}^{(k+1)}B_{t}^T\|_F^2 \\
    & \quad + \frac{\lambda_2}{2}\mathbf{I}_{\{t> 1\}}\|A_{t}^{(k+1)}B_{t}^T - A_{t-1}^{(k+1)}(B_{t-1}^{(k+1)})^T\|_F^2 \\
    & \quad + \frac{\lambda_2}{2}\mathbf{I}_{\{t< T\}}\|A_{t+1}^{(k+1)}(B_{t+1}^{(k)})^T - A_{t}^{(k+1)}B_{t}^T\|_F^2\bigg\}. \numberthis{} \label{surrogate-B}
\end{align*}

While both optimization problems are multi-target ridge regression problems, the design matrix in the problems are different across the targets due to the differences in the observed pattern $\Omega_t$. This makes it hard to solve the problem for large matrices $X_t$ because the dominant cost of solving such multi-target ridge regression problems is factorizing the design matrix. Since the design matrices are different across the targets, the algorithm has to compute a new factorization for each target. This greatly increases the computational cost of the algorithm.

To overcome this issue, we follow the idea in \cite{hastie2015matrix}, and derive an upper bound for the first term in these optimization problems that share design matrices across the targets. For example, for \eqref{surrogate-A}, we have:
\begin{equation}\label{MM}
\begin{aligned}
    \|P_{\Omega_t}(X_t - A_{t}(B_{t}^{(k)})^T)\|_F^2 &\le \|P_{\Omega_t}(X_t - A_{t}(B_{t}^{(k)})^T) + P_{\Omega_t^\perp}(A_{t}^{(k)}(B_{t}^{(k)})^T - A_{t}(B_{t}^{(k)})^T)\|_F^2 \\
    &= \|P_{\Omega_t}(X_t) + P_{\Omega_t^{\perp}} (A_{t}^{(k)}(B_{t}^{(k)})^T) - A_{t}(B_{t}^{(k)})^T\|_F^2.
\end{aligned}
\end{equation}
Equality holds when $A_{t} = A_{t}^{(k)}$. This inequality is due to the fact that $\Omega_t \cap \Omega_t^{\perp} = \emptyset$, so the squared error in both $\Omega_t$ and $\Omega_t^{\perp}$ is at least as large as the squared error for $\Omega_t$ only (the non-zero entries for the matrix on the left-hand-side is a subset of the non-zero entries for the matrix on the right-hand-side). We note that the design matrix for the multi-target ridge regression problems are all $A_t$. This allows the algorithm to factorize $A_t$ once and reuse the factorization for all the targets. This greatly reduces the computational cost of the algorithm.

Define $X^{(k)}_t=P_{\Omega_t}(X_t) + P_{\Omega_t^{\perp}} (A_{t}^{(k)}(B_{t}^{(k)})^T)$; i.e.\ $X^{(k)}_t$ is an $m\times n$ matrix, with all observed entries keeping their values as those in $X_t$ and all missing entries in $X_t$ being filled in by $A_{t}^{(k)}(B_{t}^{(k)})^T$. In light of \eqref{MM}, the objective function in \eqref{surrogate-A} is upper bounded by:
\begin{align*}
    &\tilde{Q}(A_{t}|A_{1:t-1}^{(k+1)},A_{t:T}^{(k)},B_{1:T}^{(k)}) \\
    &\triangleq \frac12\|X^{(k)}_t - A_{t}(B_{t}^{(k)})^T\|_F^2 + \frac{\lambda_1}{2}\|A_t\|_F^2 + \frac{\lambda_3}{2} \|Y_t - A_{t}(B_{t}^{(k)})^T\|_F^2\\
    &\quad + \frac{\lambda_2}{2}\mathbf{I}_{\{t> 1\}}\|A_{t}(B_{t}^{(k)})^T - A_{t-1}^{(k+1)}(B_{t-1}^{(k)})^T\|_F^2 \\
    &\quad + \frac{\lambda_2}{2}\mathbf{I}_{\{t< T\}}\|A_{t+1}^{(k)}(B_{t+1}^{(k)})^T - A_{t}(B_{t}^{(k)})^T\|_F^2. \numberthis{} \label{surrogate-A-MM}
\end{align*}

Similarly, define $X^{(k+\frac12)}_t = P_{\Omega_t}(X_t) + P_{\Omega_t^{\perp}} (A_{t}^{(k+1)}(B_{t}^{(k)})^T)$. We have the following upper bound for the objective function in \eqref{surrogate-B}:
\begin{align*}
    &\tilde{Q}(B_{t}|A_{1:T}^{(k+1)},B_{1:t-1}^{(k+1)},B_{t:T}^{(k)}) \\\
    &\triangleq \frac12\|X^{(k+\frac12)}_t - A_{t}^{(k+1)}B_{t}^T\|_F^2 + \frac{\lambda_1}{2}\|B_t\|_F^2 + \frac{\lambda_3}{2} \|Y_t - A_{t}^{(k+1)}B_{t}^T\|_F^2 \\
    & \quad + \frac{\lambda_2}{2}\mathbf{I}_{\{t> 1\}}\|A_{t}^{(k+1)}B_{t}^T - A_{t-1}^{(k+1)}(B_{t-1}^{(k+1)})^T\|_F^2 \\
    & \quad + \frac{\lambda_2}{2}\mathbf{I}_{\{t< T\}}\|A_{t+1}^{(k+1)}(B_{t+1}^{(k)})^T - A_{t}^{(k+1)}B_{t}^T\|_F^2. \numberthis{} \label{surrogate-B-MM}
\end{align*}
We note that these upper bounds have the property that 
\[
\begin{aligned}
\tilde{Q}(A_t^{(k)}\mid A_{1:t-1}^{(k+1)},A_{t:T}^{(k)},B_{1:T}^{(k)}) = Q(A_t^{(k)}\mid A_{1:t-1}^{(k+1)},A_{t:T}^{(k)},B_{1:T}^{(k)}), \\
\tilde{Q}(B_t^{(k)}\mid A_{1:T}^{(k+1)},B_{1:t-1}^{(k+1)},B_{t:T}^{(k)}) = Q(B_t^{(k)}\mid A_{1:T}^{(k+1)},B_{1:t-1}^{(k+1)},B_{t:T}^{(k)}).
\end{aligned}
\]
In other words, these upper bounds are tight at the current values of the $A_t$ and $B_t$. As we shall see, this property is crucial to the convergence of the algorithm: it ensures that minimizing the upper bound always reduces the objective value of \eqref{softALS-TSSH}.

The two upper bound objective functions can be minimized to find the updated $A_t, B_t$:
\begin{align}
    A_t^{(k+1)} = &\argmin \quad \left\{\tilde{Q}(A_{t}|A_{1:t-1}^{(k+1)},A_{t:T}^{(k)},B_{1:T}^{(k)})\right\}, \label{update-A-MM}\\
    B_t^{(k+1)} = &\argmin \quad
    \left\{\tilde{Q}(B_{t}|A_{1:T}^{(k+1)},B_{1:t-1}^{(k+1)},B_{t:T}^{(k)})\right\}. \label{update-B-MM}
\end{align}

One may notice that in order to update the $A_t, B_t$ given other matrices, we are essentially doing a multi-target ridge regression \cite{hoerl1970ridge} with shared coefficients and predictors. For example, for updating $A_t$, with $B_{t}^{(k)}$ as predictors, we are regressing on four targets: the ``filled-in" matrix $X^{(k)}_t$, the auxiliary data $Y_t$, the imputation $A_{t-1}^{(k+1)}(B_{t-1}^{(k)})^T$ at time $t-1$, and the imputation $A_{t+1}^{(k)}(B_{t+1}^{(k)})^T$ at time $t+1$. The four targets are weighed by $1, \lambda_3, \lambda_2, \lambda_2$, respectively. Define a weighted label $Z_t^{(k)}$ as:

\begin{equation}\label{weighted-label}
    Z_t^{(k)}=X_t^{(k)} + \lambda_2 \left( \mathbf{I}_{\{t> 1\}} A_{t-1}^{(k+1)}(B_{t-1}^{(k)})^T + \mathbf{I}_{\{t < T\}} A_{t+1}^{(k)}(B_{t+1}^{(k)})^T\right) + \lambda_3 Y_t.
\end{equation}

Similarly, when we update $B_t$, we define $Z_t^{(k+\frac12)}$ as:

\begin{equation}\label{weighted-label-2}
    Z_t^{(k+\frac12)} = X^{(k+\frac12)}_t + \lambda_2 \left( \mathbf{I}_{\{t> 1\}} A_{t-1}^{(k+1)}(B_{t-1}^{(k+1)})^T + \mathbf{I}_{\{t < T\}} A_{t+1}^{(k+1)}(B_{t+1}^{(k)})^T\right) + \lambda_3 Y_t.
\end{equation}

With the notation above, the updated $A_t$ in \eqref{update-A-MM} has a closed-form solution:

\begin{equation}\label{update-A}
    A_t^{(k+1)} = \left[(1+\lambda_2(\mathbf{I}_{\{t < T\}} + \mathbf{I}_{\{t > 1\}}) + \lambda_3)(B_{t}^{(k)})^TB_{t}^{(k)} + \lambda_1 \mathrm{I}\right]^{-1}Z_t^{(k)}B_{t}^{(k)}.
\end{equation}

Similarly, the updated $B_t$ in \eqref{update-B-MM} also has a closed-form solution:

\begin{equation}\label{update-B}
    B_t^{(k+1)} = \left[(1+\lambda_2(\mathbf{I}_{\{t < T\}} + \mathbf{I}_{\{t > 1\}}) + \lambda_3)(A_{t}^{(k+1)})^TA_{t}^{(k+1)} + \lambda_1 \mathrm{I}\right]^{-1}(Z_t^{(k+\frac12)})^TA_{t}^{(k+1)}.
\end{equation}
The two closed-form solutions for $A_t^{(k+1)},B_t^{(k+1)}$ are shrinkage estimators (\cite{efron1976multivariate,lehmann2006theory}). Compared to the original softImpute-ALS method, the shrinkage towards zero (which is dictated by the regularization term for $A$ and $B$ matrices) is not as large if given the same $\lambda_1$ and non-zero $\lambda_2,\lambda_3$ values. Therefore, one can expect imputation matrices with higher rank out of the algorithm with either temporal smoothing ($\lambda_2 > 0$) or auxiliary data ($\lambda_3 > 0$). {In other words, the presence of any non-zero $\lambda_2$ and $\lambda_3$ make the shrinkage effect introduced by $\lambda_1$ weaker. One can easily lower the value of $\lambda_1$ to make the shrinkage of $A$ and $B$ weaker without introducing $\lambda_2$ or $\lambda_3$, but the softImpute-ALS algorithm itself (when $\lambda_2=\lambda_3=0$) cannot guarantee the temporal and spatial smoothness and may over-fit the observed regions.} This shrinkage formula exemplifies the role that each component plays: the regularization, the auxiliary data, and the temporal smoothing. And the roles that the tuning parameters $\lambda_1,\lambda_2,\lambda_3$ play are also clear from this shrinkage formula: the relative values of these three tuning parameters determine the extent of sparsity of the imputed matrices, the extent of the temporal smoothing, and the weight of the auxiliary data. 

The two shrinkage formulas for $A_t$ and $B_t$ showcase a mutual normalization phenomenon: the extent of the shrinkage in estimating $A_t$ depends on the current values of $B_t$ and vice versa. This results from the matrix factorization assumption and is quite interesting from the perspective of empirical Bayes estimators. One can easily formulate the optimization problem in \eqref{softALS-TSSH} as solving the maximum-a-posterior estimate of a Bayesian model with latent factored Markov structures. In that case, $\lambda_1^{-1}$, $\lambda_2^{-1}$, and $\lambda_3^{-1}$ serve as prior variance for $A_t, B_t$, the variance of the Gaussian Markovian latent structure, and the variance of the auxiliary data. We omit further details of this here to not deviate too much from the main results of the current paper. 


Additionally, compared with existing matrix completion methods, we use weighted labels $Z_t^{(k)}, Z_t^{(k+\frac12)}$ that include imputations of neighboring time points on top of the original data when learning the imputation. A related work \cite{wang2014low} also accounts for temporal smoothness in matrix completion, but they use alternating direction method (ADM) \cite{chen2012matrix}. Also, our weighted label incorporates the auxiliary data that are particularly helpful for imputing scientific images with large patch missingness.

In a single iteration, we update $A_t$ using \eqref{update-A} for each $t$ and update $B_t$ using \eqref{update-B} for each $t$. The algorithm terminates when all imputations $A_t^{(k)}(B_t^{(k)})^T$ converge. To verify convergence, we calculate the relative change of the Frobenius norm:
\begin{equation}\label{frob_norm_change}
    \nabla F_t^{(k)} = \frac{\|A_t^{(k+1)}(B_t^{(k+1)})^T - A_t^{(k)}(B_t^{(k)})^T\|_F^2}{\|A_t^{(k)}(B_t^{(k)})^T\|_F^2}.
\end{equation}
The termination rule that we set for the iterative algorithm is that the algorithm stops at iteration $k$ if $\max\{\nabla F_1^{(k)}, \nabla F_2^{(k)}, \dots, \nabla F_T^{(k)}\}$ is smaller than a pre-specified threshold.

The full algorithm uses a singular value decomposition (SVD) form to express $A_t, B_t$: $A_t = U_{t}D_t, B_t = V_{t}D_t$, following the original softImpute algorithm. This guarantees $\|A_{t}B_{t}^T\|_{*} = \frac12 (\|A_t\|^{2}_{F} + \|B_t\|^{2}_{F})$, which means that the algorithm is equally penalizing the trace norm of the imputation. We present the full algorithm below.

\begin{algorithm}[H]
\label{algorithm}
\caption{softImpute-ALS with Temporal Smoothing and Auxiliary Data}
    \begin{algorithmic}[1]
    \INPUT $m\times n$ Sparse data $X_1, X_2, \dots, X_T$, $m\times n$ auxiliary data $Y_1, Y_2, \dots, Y_T$, operating rank $r$. Maximum iteration $\mathrm{K}$ and convergence threshold $\tau$.
    \OUTPUT Imputation of sparse data $A_1B_1^T, A_2B_2^T, \dots, A_TB_T^T$.
    \STATE \textbf{Initialization:} For $1\leq t \leq T$, $A_t^{(1)}=U_tD_t, B_t^{(1)}=V_tD_t$, where $U_t, V_t$ are $m\times r, n\times r$ randomly chosen matrix with orthogonal columns. $D_t$ is $\mathrm{I}_{r\times r}$
    \STATE \textbf{Update A:}
    \FOR{$t=1:T$}
    \Indent
    \STATE a. Let $X^{(k)}_t=P_{\Omega_t}(X_t) + P_{\Omega_t^{\perp}} (A_{t}^{(k)}(B_{t}^{(k)})^T)$, which is the ``filled-in" version of $X_t$
    \STATE b. Let $Z_t^{(k)}$ be the weighted label in equation \eqref{weighted-label}
    \STATE c. $A_t^{(k+1)}$ is updated as equation \eqref{update-A}
    \EndIndent
    \ENDFOR
    \STATE \textbf{Update B:} For every $t$, repeat a,b,c steps above, with $X^{(k)}_t, Z_t^{(k)}$ being replace by $X^{(k+\frac12)}_t, Z_t^{(k+\frac12)}$. $B_t^{(k+1)}$ is calculated following equation \eqref{update-B}
    \STATE Repeat updating $A_{1:T}$ and $B_{1:T}$ until convergence. The algorithm converges when $\max\{\nabla F_1^{(k)}, \nabla F_2^{(k)}, \dots, \nabla F_T^{(k)}\} < \tau$, with $\nabla F_t^{(k)}$ defined in \eqref{frob_norm_change}.
    \STATE For any $t$, denote the final output as $A_t^{*}, B_t^{*}$. Let $X_t^*=P_{\Omega_t}(X_t) + P_{\Omega_t^{\perp}} (A_{t}^{*}(B_{t}^{*})^T)$. 
    \Indent
    \STATE Do SVD for $A_t^{*}(B_t^{*})^T=U_t^{*}(D_t^{*})^2(V_t^{*})^T$
    \STATE Define $M_t=X_t^{*}V_t^{*}$ and do SVD for $M_t=\tilde{U}_t\tilde{D}_tR_t^T$.
    \STATE Do soft-thresholding on $\tilde{D}_t$: $\tilde{D}_{t,\lambda_1} = \mathbf{diag}\left[(\sigma_1-\lambda_1)_+,(\sigma_2-\lambda_1)_+,\dots,(\sigma_r-\lambda_1)_+\right]$
    \STATE Output imputation for time $t$ as $\tilde{U}_t\tilde{D}_{t,\lambda_1}(V_t^{*}R_t)^T$
    \EndIndent
    \end{algorithmic}
\end{algorithm}

The last few steps in the algorithm adopts the idea of softImpute-ALS in \cite{hastie2015matrix} to ensure that the output imputation matrix is sparse and has zero singular values.

Before moving on to the discussion of the theoretical property of the algorithm, we want to note that there are three hyper-parameters in our algorithm: $\lambda_1,\lambda_2,\lambda_3$, and potentially more hyper-parameters in the process of generating the auxiliary data. It would be computationally demanding if one wants to search for the best parameters via grid search and cross-validation. Based on our empirical experience, our method provides stable imputation result for a wide range of hyper-parameters. Therefore, we recommend our readers to do a sequential search: search for the best $\lambda_1$ first, and then search for the best $\lambda_2$ and $\lambda_3$ in parallel. One can determine the initial range of these parameters using a relatively small dataset (i.e. with only a few frames) and it typically generalizes well to the entire dataset. In the following texts, we use this sequential search method and pick the best hyper-parameters based on the performance on a left-alone validation set. Specifically for TEC map reconstruction, we recommend $\lambda_1=0.8\sim1.0, \lambda_2=0.2\sim0.3, \lambda_3=0.02\sim0.03$.

Another tuning parameter is $r$, which is the other dimension of the factor matrices $A$ and $B$, is set to $\min\{m,n\}$ in our algorithm. Basically, we allow for the maximum possible rank for the imputed map. In other applications, if one has prior knowledge about the low-rank nature of the data, one can set an initial low rank structure. Even though we do not take advantage of this in the TEC imputation task, we still make it a modelling possibility for other applications.

\subsection{Theoretical Properties of the Algorithm}\label{theory-section}

In this section, we provide theoretical results for the convergence rate of the algorithm and show that the algorithm converges to a stationary point of the problem defined in Equation \eqref{softALS-TSSH}. Proofs, following those in \cite{hastie2015matrix}, are included in the Supplemental Materials \cite{Supplement}.

Recall that the objective function to minimize is defined in \eqref{softALS-TSSH}, which is $F(A_{1:T},B_{1:T})$. We approach this optimization problem by the majorization-minimization and alternating least squares (ALS) methods. The first theoretical result, Theorem~\ref{descent-theorem}, indicates that each update on $A_t$ and $B_t$ does not not increase the objective function value.

Let $\{A_{1:T}^{(k)}, B_{1:T}^{(k)}\}_{k\geq 1}$ be the sequence of $A_{1:T},B_{1:T}$ generated through the iterations of Algorithm 1, where $A_{1:T}^{(k)},B_{1:T}^{(k)}$ denotes the matrices at iteration $k$. Define the descent of objective function value at iteration $k$ as $\Delta_k = F(A_{1:T}^{(k)},B_{1:T}^{(k)}) - F(A_{1:T}^{(k+1)},B_{1:T}^{(k+1)})$.

\begin{theorem}\label{descent-theorem}
Then the value of the objective function is non-increasing, i.e.,
\begin{equation*}
    F(A_{1:T}^{(k)}, B_{1:T}^{(k)}) \geq F(A_{1:T}^{(k+1)}, B_{1:T}^{(k)}) \geq F(A_{1:T}^{(k+1)}, B_{1:T}^{(k+1)}),
\end{equation*}
thus $\Delta_k\geq 0$, for all $k\geq 1$. 
\end{theorem}
Proof of this theorem is in Section A.1 of the Supplement Material \cite{Supplement}.

Theorem \ref{descent-theorem} indicates that each iteration is making the objective function smaller. Given that each matrix $A_t,B_t$ is updated by a ridge regression, we show that, in Theorem~\ref{descent-LB}, the decrease of objective function value in each iteration has a lower bound.
\begin{theorem}\label{descent-LB}
We have the following lower bound for $\Delta_k$:
\begin{align*}
    \Delta_k & \geq \frac{\lambda_1}{2}\sum_{t=1}^T \left(\|A_t^{(k)}-A_t^{(k+1)}\|^2 + \|B_t^{(k)}-B_t^{(k+1)}\|^2 \right)\\
    & + \frac{1}{2} \sum_{t=1}^T (1+\lambda_2(1+\mathbf{I}_{\{2\leq t\leq T-1\}})+\lambda_3) \left[\delta_{k,t}\right], \numberthis{}
\end{align*}
where $\delta_{k, t} = \|(A_t^{(k)}-A_t^{(k+1)})(B_t^{(k)})^T\|^2 + \|A_t^{(k+1)}(B_t^{(k)}-B_t^{(k+1)})^T\|^2$.
\end{theorem}
The proof if this theorem is in Section A.2 of the Supplement Material \cite{Supplement}.

This result gives a lower bound for the descent of the objective function value at iteration step $k$. We look at the first term. With a non-zero $\lambda_1$, as long as there exists one $A_t$ or $B_t$ that has a different value/entry before and after the update, the lower bound is greater than zero. Thus $\Delta_k$ may be interpreted as a measure of the optimality of $\{(A_t^{(k)},B_t^{(k)})\}_{t=1}^T$. Given the result in theorem \ref{descent-theorem} and \ref{descent-LB}, the sequence of objective function values is a bounded, strictly monotonic sequence, thus having a finite limit (may not be unique, depending on the initialization), denoted as $f^{\infty}$. The following result, Theorem~\ref{conv-rate-theorem}, gives the convergence rate for our algorithm.
\begin{theorem}\label{conv-rate-theorem}
Let the limit of the objective function $F(A_{1:T}^{(k)},B_{1:T}^{(k)})$ be $f^{\infty}$, we have:
\begin{equation}\label{conv-rate}
    \min_{1\leq k\leq K} \Delta_k \leq \frac{F(A^{(1)}_{1:T},B^{(1)}_{1:T}) - f^{\infty}}{K},
\end{equation}
where K is the total number of iterations. Additionally, assume that there exists positive constants $l^L$ and $l^U$ such that $l^{L}\mathrm{I}\leq(A_t^{(k)})^{T}A_t^{(k)}\leq l^{U}\mathrm{I}$, $l^{L}\mathrm{I}\leq(B_t^{(k)})^{T}B_t^{(k)}\leq l^{U}\mathrm{I}$ for all $t,k$, then we have:
\begin{align*}\label{conv-1}
    \min_{1\leq k\leq K} \left\{\sum_{t=1}^T \left(\|A_t^{(k)}-A_t^{(k+1)}\|^2 + \|B_t^{(k)}-B_t^{(k+1)}\|^2 \right)\right\} \\
    \leq \frac{2}{(1+\lambda_2+\lambda_3)l^{L}+\lambda_1} \left(\frac{F(A^{(1)}_{1:T},B^{(1)}_{1:T}) - f^{\infty}}{K}\right), \numberthis{}
\end{align*}
and
\begin{align*}\label{conv-2}
    \min_{1\leq k\leq K} \left\{\sum_{t=1}^T  \left(\|(A_t^{(k)}-A_t^{(k+1)})(B_t^{(k)})^T\|^2 + \|A_t^{(k+1)}(B_t^{(k)}-B_t^{(k+1)})^T\|^2\right)\right\} \\
    \leq \frac{2l^U}{l^{U}(1+\lambda_2+\lambda_3)+\lambda_1} \left(\frac{F(A^{(1)}_{1:T},B^{(1)}_{1:T}) - f^{\infty}}{K}\right). \numberthis{}
\end{align*}
\end{theorem}
The proofs are very similar to that of Theorem 4 and Corollary 1 of \cite{hastie2015matrix}. Thus we only briefly describe the proof in Section A.3 of the Supplement Material \cite{Supplement}.

Recall $\Delta_k$ is a measure of the optimality of $\{(A_t^{(k)},B_t^{(k)})\}_{t=1}^T$. Theorem \ref{conv-rate-theorem} implies this measure of optimality converges at a $\frac1K$-rate. We note that this is the same rate of convergence as softImpute-ALS, which is $\mathrm{O}(1/K)$. There are two additional notions of convergence, namely \eqref{conv-1} and \eqref{conv-2}, that show the role of $\lambda_1, \lambda_2, \lambda_3$ in the convergence rate. Generally, fewer iterations are needed given larger $\lambda_1, \lambda_2, \lambda_3$. 

The last result follows the theorem 5 in \cite{hastie2015matrix}, which states that the limit point of the sequence $\{A_{1:T}^{(k)},B_{1:T}^{(k)}\}_{k\geq 1}$, denoted by $A_{1:T}^{*},B_{1:T}^{*}$, is a stationary point of problem \eqref{softALS-TSSH}, so we state Theorem~\ref{thm:final} here without a proof.
\begin{theorem}\label{thm:final}
Let $\{A_{1:T}^{(k)}, B_{1:T}^{(k)}\}_{k\geq 1}$ be the sequence of $A_{1:T},B_{1:T}$ generated throughout the iterations of Algorithm 1. For $\lambda_1>0$, the limit point of the sequence $A_{1:T}^{*},B_{1:T}^{*}$ is a stationary point of problem \eqref{softALS-TSSH} in the sense that:
\begin{align*}
    A_t^{*} = &\argmin \quad \left\{\tilde{Q}(A_{t}|A_{1:T}^{*},B_{1:T}^{*})\right\}, \\
    B_t^{*} = &\argmin \quad
    \left\{\tilde{Q}(B_{t}|A_{1:T}^{*},B_{1:T}^{*})\right\}. 
\end{align*}
\end{theorem}

Just as in the softImpute paper~\cite{hastie2015matrix}, there is no theoretical guarantee that the sequence of matrices generated by alternating least squares converges to the global minimum of the optimization problem \eqref{softALS-TSSH}. What we have proved and stated in this section is that the algorithm improves the objective function at each iteration, converges to a stationary point, and the three tuning parameters have their roles in the rate of convergence. In practice, our algorithm performs well on the video/matrix imputation task.

Since we use cyclic least square method on updating $2T$ matrices, the computational cost of the algorithm is about $T$ times of the cost of softImpute-ALS which updates $2$ matrices per iteration. But our approach imputes $T$ matrices simultaneously, the computational cost is thus similar to softImpute-ALS per iteration. In practice, the algorithm converges in fewer iterations when including the temporal smoothing and auxiliary data penalty, so the algorithm is indeed less time-consuming than softImpute-ALS.

In the next section, we use a carefully designed simulation study to compare our method with the baseline softImpute-ALS in terms of the accuracy of the imputation. Then we present the imputation results for TEC maps using our method and demonstrate how our method resolves the two issues with softImpute-ALS when imputing TEC videos described in Section~\ref{subsec:litreview}.

\section{Numerical Studies}
To compare our proposed method with the softImpute-ALS and the state-of-the-art of the scientific field (Spherical Harmonics), we use the TEC maps provided by the International GNSS Service (IGS). The advantage of using this data is that there are no missing values. Thus we can create missing values artificially and still know the original values of the missing entries. We pick the 1-day IGS data on Sept-08, 2017, which contains 96 matrices at a 15-min resolution. Each TEC map is of size $71\times 73$, so the video is of size $71\times 73\times 96$. We resize each TEC map to $181\times 361$ with bi-linear interpolation to match the size of the TEC map of the madrigal database, which is the observed TEC maps with missing values that we want to impute. 

\begin{figure}[h]
    \centering
    \includegraphics[width=0.98\textwidth]{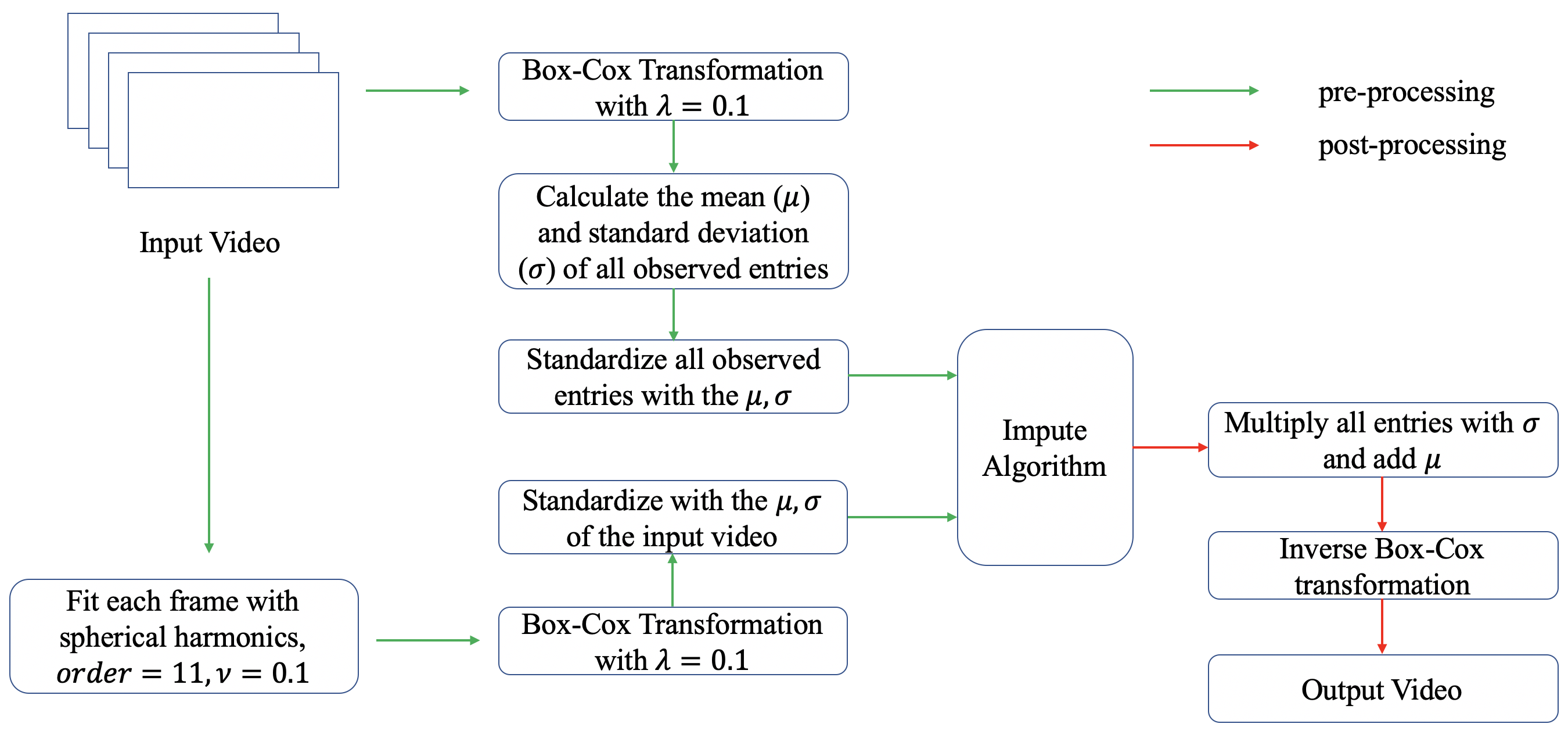}
    \caption{Data analysis pipeline: video imputation. Input video contains missing values. Spherical Harmonics is fitted on the input video with a carefully chosen order and penalty weight $v$ to optimize its performance. Standardization is done for all observed pixels in both data. To obtain output video, we invert both standardization and the Box-Cox transformation to make sure the input and output videos have comparable magnitudes.}
    \label{fig:pipeline}
\end{figure}

In this section, we describe 4 different ways of ``creating" missing values in the IGS data. After creating missingness, we then generate auxiliary data and apply our imputation algorithms. In Figure \ref{fig:pipeline}, we illustrate the data pipeline from any input video (with missing values) to the output video (imputed  full videos). This pipeline is applied to all numerical analyses and empirical analyses in the paper.

We apply the Box-Cox transformation \cite{box1964analysis} on each observed pixel of the input video and auxiliary data (i.e. spherical harmonics fitted data) to make the data more like normally-distributed. Pixel-wisely, the Box-Cox transformation is doing $y' = \frac{y^\lambda - 1}{\lambda}$ for any pixel intensity $y$. This could make the imputation more robust to extreme values.

Before going into details of the design and results of numerical analysis, we first give a brief overview of the spherical harmonics fitting method, which is the method we use to generate auxiliary data from the input video.

\subsection{A Brief Note on Spherical Harmonics Fitting}\label{subsec:SH}

Known as the angular portion of the solutions to the Laplace's equation in spherical coordinates, spherical harmonic functions define a complete orthogonal basis and therefore can approximate a sufficiently smooth surface function in the form:
$$f(\theta,\phi) \approx \sum^{l_{max}}_{l=0}\sum^l_{m=-l} a_l^mY_l^m(\theta,\phi),$$
where $\theta$ and $\phi$ are the elevation and azimuth angles in the spherical coordinates. $Y_l^m(\theta,\phi)$ denotes a spherical harmonic function with degree $m$ and order $l$ ($|m|\leq l$), and $a_l^m$ is the corresponding coefficient. Similar to the Fourier series, when the maximum order $l_{max} \to \infty$, the expansion becomes an exact representation of the function $f(\theta,\phi)$.

\begin{figure}[h]
    \centering
    \includegraphics[width=0.9\textwidth]{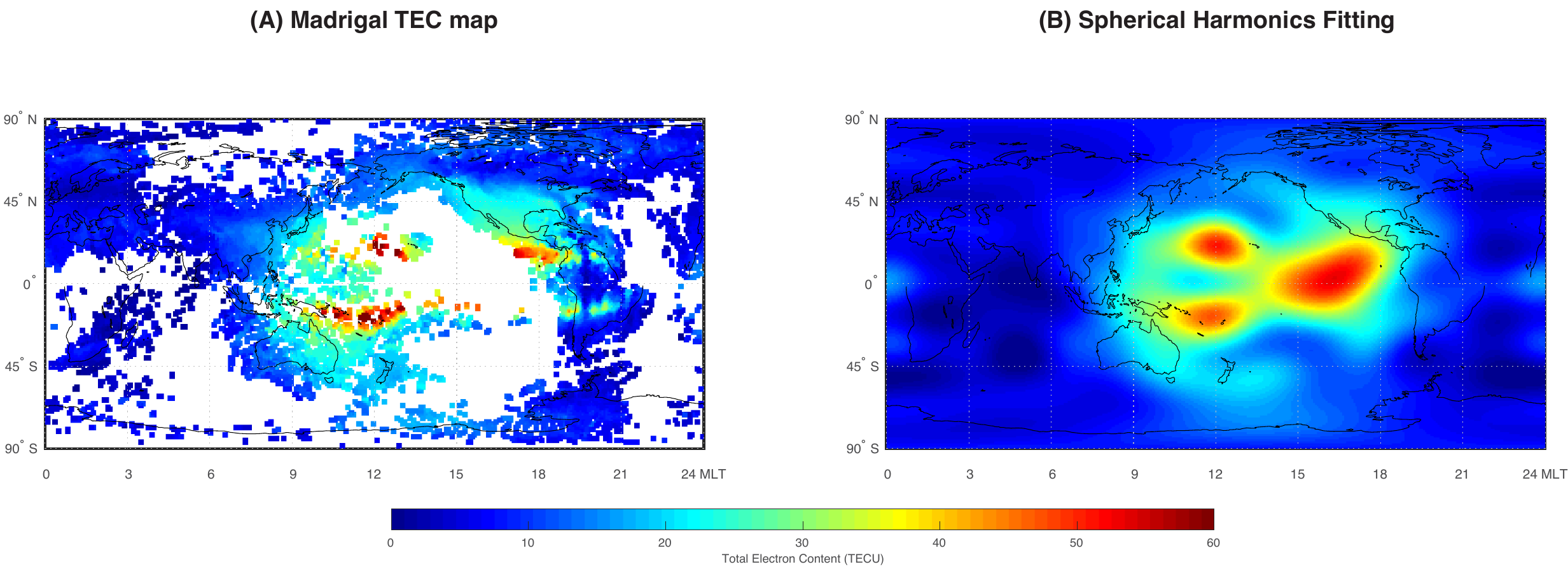}
    \caption{(A) Madrigal TEC map with missing data and (B) complete TEC map approximated by the spherical harmonics expansion.}
    \label{fig:sh_fit}
\end{figure}

By viewing the global TEC distribution at a given time as a function of latitude and longitude, we can use the spherical harmonic expansion as an approximation to the complete TEC map. Each single measurement of TEC at location $(\theta_i,\phi_i)$ can provide a linear equation $f(\theta_i,\phi_i) = \text{TEC}_i = \sum^{l_{max}}_{l=0}\sum^l_{m=-l} a_l^mY_l^m(\theta_i,\phi_i)$, and by solving a system of linear equations formed using all the available measurements on a TEC map, a set of coefficients $a_l^m$ can be obtained for a given $l_{max}$, and the resulting expansion is the least squares approximation of the global TEC map (Figure \ref{fig:sh_fit}). To avoid high-frequency artifacts and negative TEC values in the spherical harmonic fitting result, Tikhonov regularization and inequality constrains \cite{Zhang2013} were applied to solve the least squares problem.

As one can see, the fitted map provides some reasonable ``observations" in the large missing patches, which could potentially improve the imputation in the oceanic areas. In our analysis, we fit any TEC map with missing values using Spherical Harmonics to get the auxiliary data for TEC map, and we choose the order $l_{max}=11$ and the weight of Tikhonov regularization $v = 0.1$ based on cross-validation.

\subsection{Description of Design of Numerical Experiments}

\begin{figure}[h]
    \centering
    \includegraphics[width=0.98\textwidth]{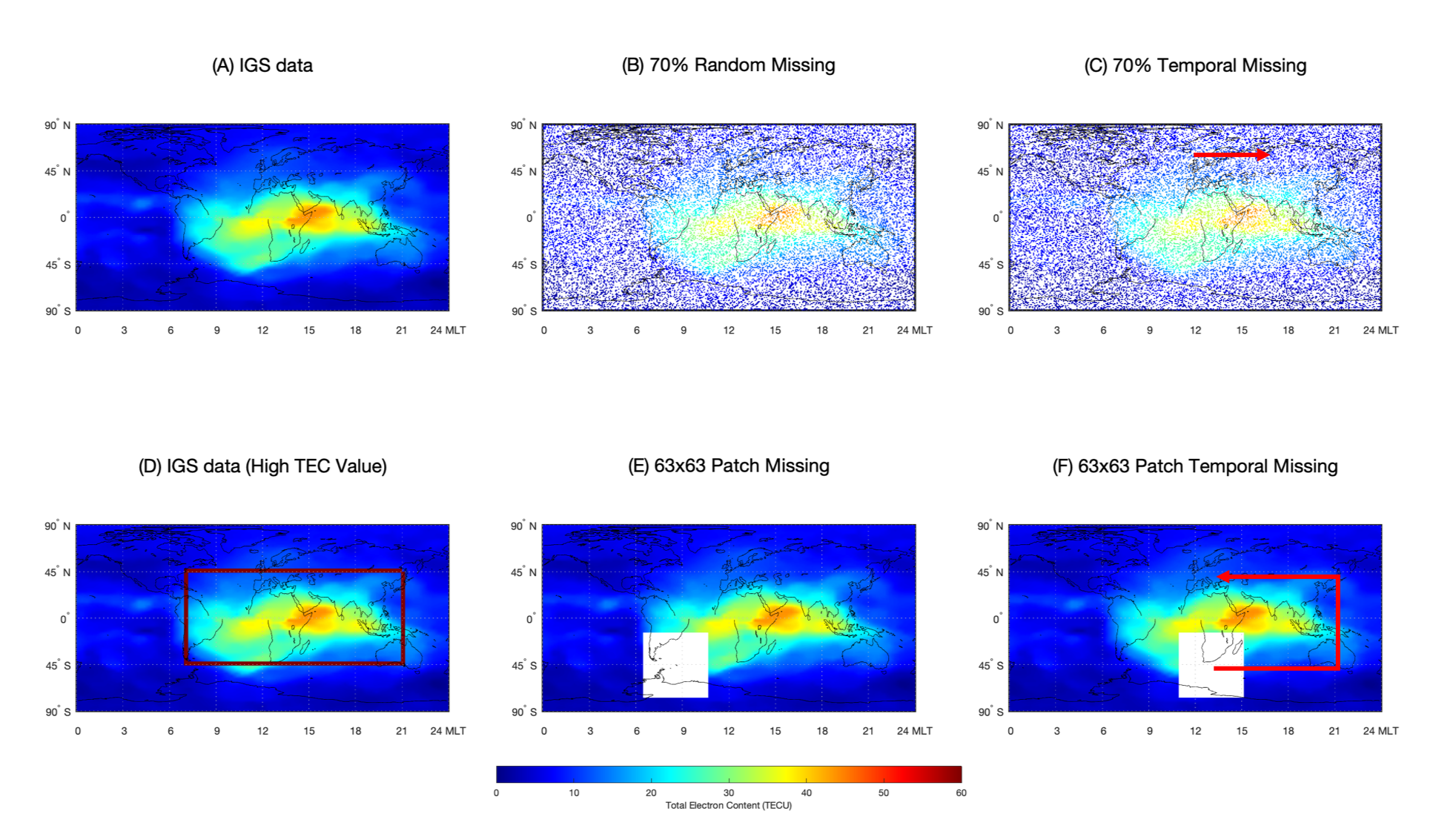}
    \caption{Four missingness patterns, where white pixels denote missing values. (A) $181\times 361$ TEC map (IGS data) at 2017-09-08 11:57:30 UT. (B) Pattern 1: Random missingness. (C) Pattern 2: Temporal missingness. (D) $181\times 361$ TEC map, with a bounding box around region $\left[45^{\circ} \mathrm{N}, 45^{\circ} \mathrm{S}\right]\times [7 \mbox{ MLT},21 \mbox{ MLT}]$ with high TEC values. (E) Pattern 3: Random Patch missingness. (F) Pattern 4: Temporal patch missingness.}
    \label{fig:four_missing}
\end{figure}

Given the resized $181\times 361\times 96$ TEC map from IGS dataset, we introduce four data missingness patterns in each of the $96$ matrices for numerical analysis. An example of the original TEC map with various missingness patterns is shown in Figure \ref{fig:four_missing}.

\begin{enumerate}
    \item Random missingness (sub-figure B): for each matrix, randomly drop $30\%/50\%/70\%$ of the pixels.
    \item Temporal missingness (sub-figure C): for the first matrix, randomly drop $30\%/50\%/70\%$ of pixels, and let the missing mask move 6 columns horizontally (direction shown as the red arrow).
    \item Random patch missingness (sub-figure E): for each frame, randomly pick a center on a fixed bounding box around high TEC value region (sub-figure D) and create a $27\times 27$ or $45\times 45$ or $ 63\times 63$ patch as missing.
    \item Temporal patch missingness (sub-figure F): similar to patch missingness, but the center of the $27\times 27/45\times 45/ 63\times 63$ patch moves along the bounding box at the speed of 6 columns(rows) per matrix (anti-clockwise as shown by the red arrow).
\end{enumerate}


Random/Temporal missingness are meant to emulate a random/auto-correlated, scattered data missingness pattern as can be observed in some local regions of the TEC map. Random/Temporal patch missingness are simulating the large patch missingness in the TEC map. The difference is that temporal patch missingness has largely overlapping patches in temporally adjacent matrices, further restricting the information available from the temporal dimension. In the madrigal database, the missing pattern is a mixture of scattered-missing and patch-missing, and mainly patch-missing. In the original context where softImpute is applied, such as the Netflix competition, scattered missingness is more common.

\subsection{Results from Numerical Studies}

After applying each data missingness pattern to the IGS data, we can run softImpute-ALS and our proposed algorithm with various choices of hyper-parameters to impute the data. To evaluate model performance on imputing the missing values, we use the relative squared error (RSE) (as used in \cite{liu2012tensor,wang2014low}):
\begin{equation*}
{\rm RSE}(X_t, X_t^*,\Omega_t)=\frac{\|\mathrm{P}_{\Omega_{t}^{\perp}}(X^{*}_{t} - X_{t})\|_F}{\|\mathrm{P}_{\Omega_{t}^{\perp}}(X_{t})\|_F}, 
\end{equation*}
where $X_t$ is the fully-observed IGS data,  $\Omega_{t}$, $\mathrm{P}_{\Omega_{t}^{\perp}}(.)$ follow the definition in section \ref{method-section},  $X^{*}_t$ is the imputation of $\mathrm{P}_{\Omega_{t}}(X_t)$ and $\|.\|_F$ is the Frobenius norm. The RSE is measuring the imputation performance on the missing pixels. Throughout the rest of this paper, we report the RSE in the unit of percentages (\%).

For each data missingness pattern and each level of missingness, we fit four models to impute the matrices, where each model is a variant of the optimization problem in equation \eqref{softALS-TSSH}. Here is a list of the four models that we compare and their short-hand notations.

\begin{enumerate}
    \item \textbf{soft}: softImpute as in \cite{hastie2015matrix}: $\lambda_1=0.9, \lambda_2=0, \lambda_3=0$.
    \item \textbf{TS}: softImpute + temporal smoothing: $\lambda_1=0.9, \lambda_2=0.05, \lambda_3=0$.
    \item \textbf{SH}: softImpute + auxiliary data based on spherical harmonics: $\lambda_1=0.9, \lambda_2=0,\lambda_3=0.01$.
    \item \textbf{TS+SH}: softImpute + temporal smoothing + auxiliary data based on spherical harmonics: $\lambda_1=0.9, \lambda_2=0.05,\lambda_3=0.01$.
\end{enumerate}

The tuning parameters above are chosen for demonstration purposes instead of being selected from an extensive grid search as we do for real data. The reason why we set all $\lambda_1=0.9$ is because as one will see in the empirical section, this value works well for the baseline softImpute method when imputing TEC maps in general. Thus our choice of the tuning parameters is actually maximizing the performance of the softImpute algorithm. And we will show that even in this setting, our newly proposed algorithms outperform in majority of cases. And we fix $\lambda_2=0.05$ and $\lambda_3=0.01$, which are not the parameters that give the best performances. We pick these parameters just to show that the method can produce decent results even when one does not tune the hyper-parameters.

For each of the 96 matrices of the IGS data, we calculate the RSE on the missing pixels. Suppose that the RSE of imputing matrix at time $t$ for the four models are $\mbox{RSE}_t^{(\mbox{soft})},\mbox{RSE}_t^{(\mbox{TS})}$, $\mbox{RSE}_t^{(\mbox{SH})}, \mbox{RSE}_t^{(\mbox{TS+SH})}$, respectively. We define the test set RSE margin over softImpute for the three variants of our model as: $\Delta\mbox{RSE}^{(k)}_t = \mbox{RSE}_t^{(\mbox{soft})} - \mbox{RSE}_t^{(k)}, k \in \{\mbox{TS,SH,TS+SH}\}$. Then $\Delta\mbox{RSE}^{(k)}_t > 0$ means that model $k$ performs, on average, better than the softImpute method on imputing the missing value of matrix at time $t$. One can easily test if any model outperforms the baseline softImpute method in all 96 matrices with a simple parametric test, which is trivial to show thus we omit the detailed testing results in this paper.

\begin{figure}[h]
    \centering
    \includegraphics[width=0.8\textwidth]{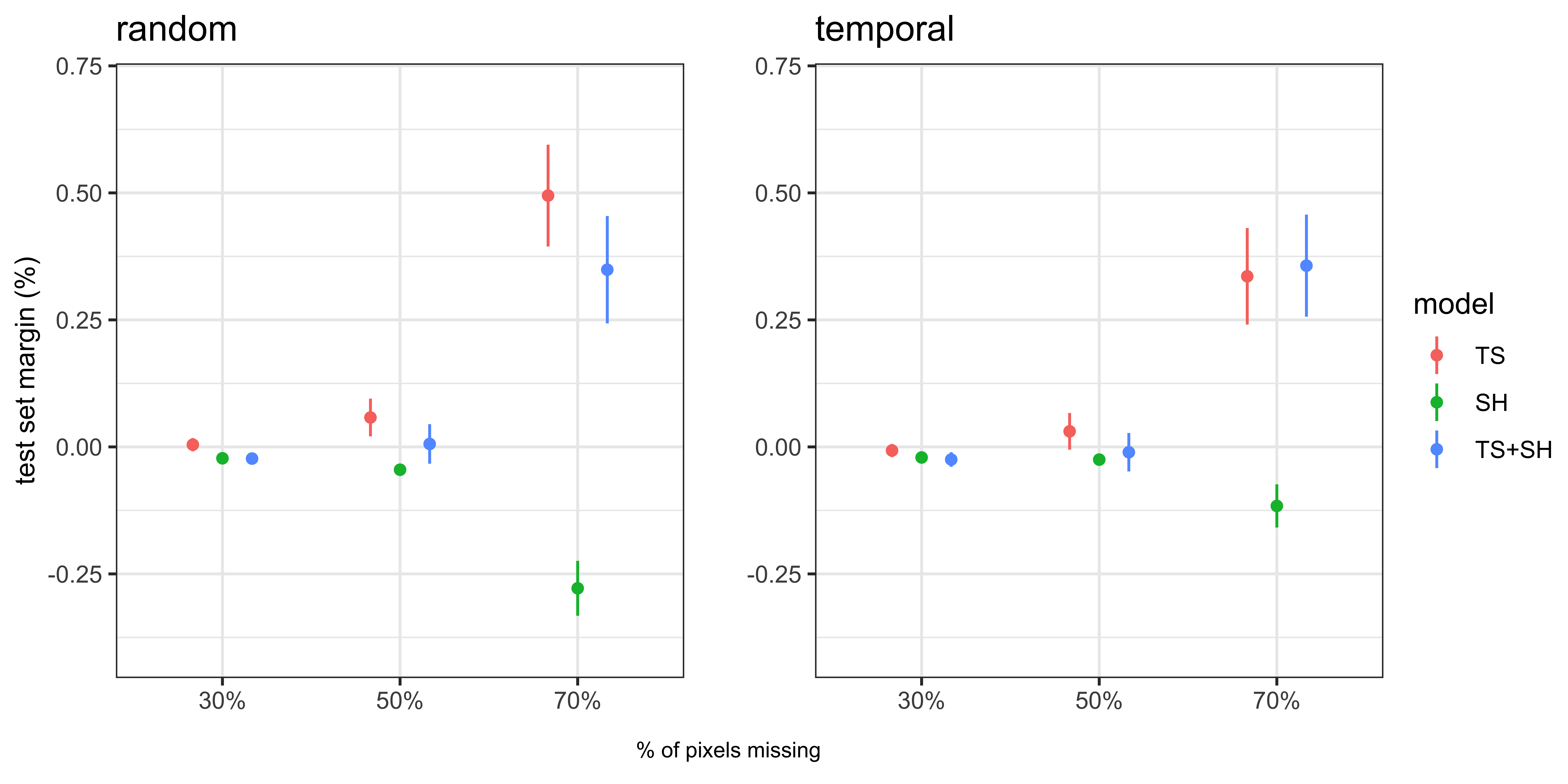}
    \caption{Numerical Analysis: random missing and temporal missing results. Three variants of our method are considered: TS, SH, TS+SH, detailed explanation included in the main text. The scatter points show the average test set RSE margin over baseline softImpute method, positive means performance better than softImpute. Error bar gives the $95\%$ confidence interval.}
    \label{fig:nonpatch-result}
\end{figure}

In Figure \ref{fig:nonpatch-result}, we report the average margin over softImpute across all 96 matrices for three variants of our model under the random missing and temporal missing scenarios. 

As one can see, as the level of missingness becomes higher, the temporal smoothing (TS) model and the full model (TS+SH) start to build a positive margin while the spherical harmonics (SH) model performs worse than the softImpute. However, the corresponding margins, regardless of being positive/negative, are all close to zero. This is mostly because under scattered missingness, the dominant information source for imputing missing entries are from neighbors that are spatially close thus temporal smoothing cannot help much. Spherical harmonics, on the other hand, has over-smoothed the data and gives some misleading information on the missing entries but the overfitting is not very severe.

\begin{figure}[h]
    \centering
    \includegraphics[width=0.8\textwidth]{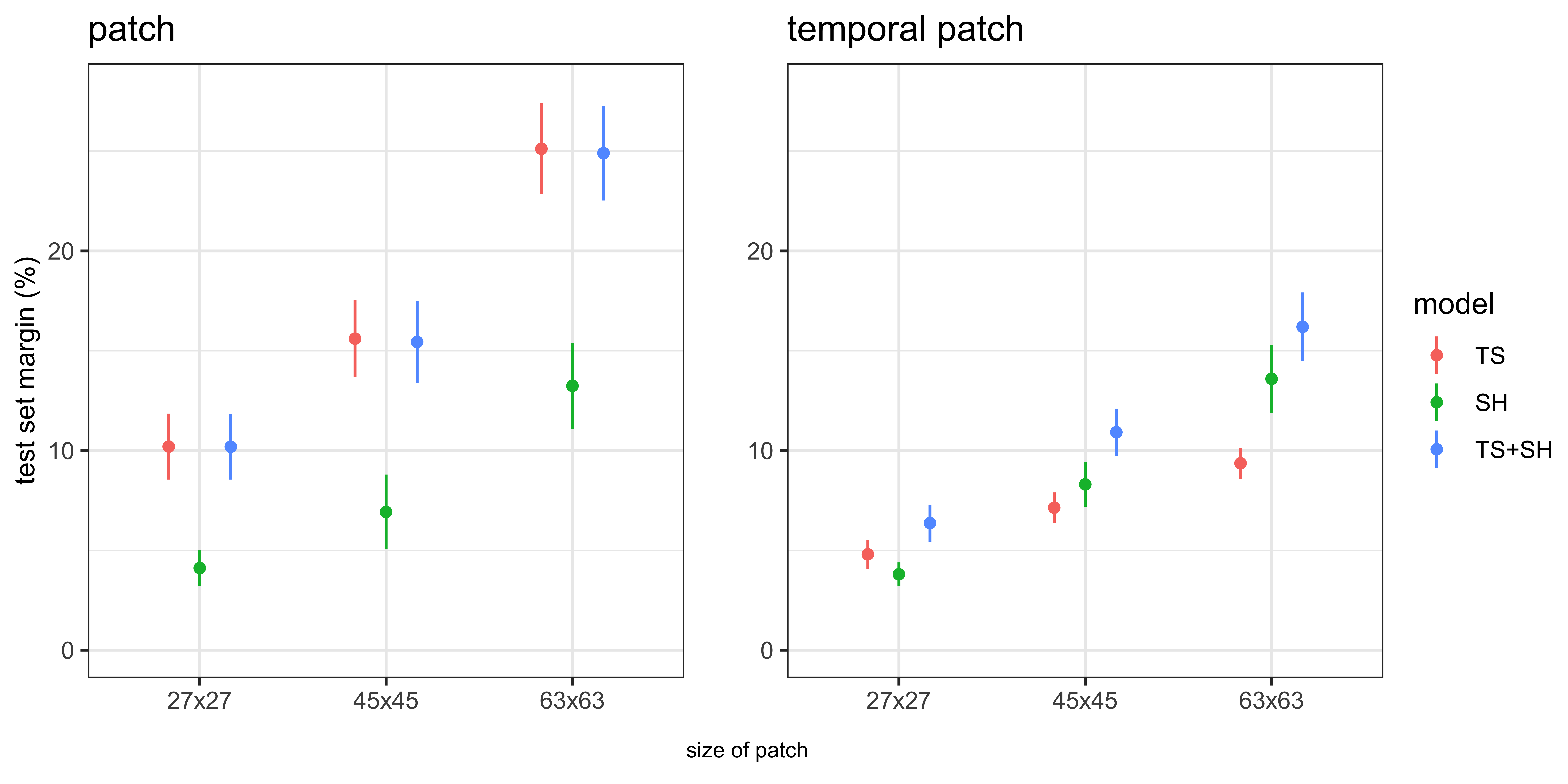}
    \caption{Numerical Analysis: random patch missing and temporal patch missing results. Three variants of our method are considered: TS, SH, TS+SH, detailed explanation included in the main text. The scatter points show the average test set RSE margin over baseline softImpute method, positive means performance better than softImpute. Error bar gives the $95\%$ confidence interval.}
    \label{fig:patch-result}
\end{figure}

The more interesting scenario is when we introduce the patch missingness, which resembles the real data more. In Figure \ref{fig:patch-result}, we make a similar plot. We observe a drastic differences between the scales. With patch missingness, all three models perform significantly better than softImpute, with the smallest margin being around $4\%$. Similarly, the higher the level of missingness, the greater the margin.

What's even more intriguing is the difference between random patch missing and temporal patch missing. When patches are randomly missing, the missing patch in the matrix at $t$ is very likely to be fully or partially observed in the matrices at $t-1$ and $t+1$, thus the temporal smoothing (TS) would significantly help on imputing the missing patches at $t$. However, when the location of the missing patch is highly auto-correlated, the missing patches in the matrices at $t-1, t$ and $t+1$ will have a large overlap\footnote{In our design, each patch moves 6 columns/rows anti-clockwise per matrix. So for patch of any size, only 6 columns or rows of the missing patch can be observed in temporally adjacent matrices.}, making temporal smoothing not as significant for improving the imputation as opposed to the previous scenario. Spherical harmonics, instead, can provide extra information over the whole missing patch, leading to more stable performance gains.

\begin{figure}[h]
    \centering
    \includegraphics[width=0.98\textwidth]{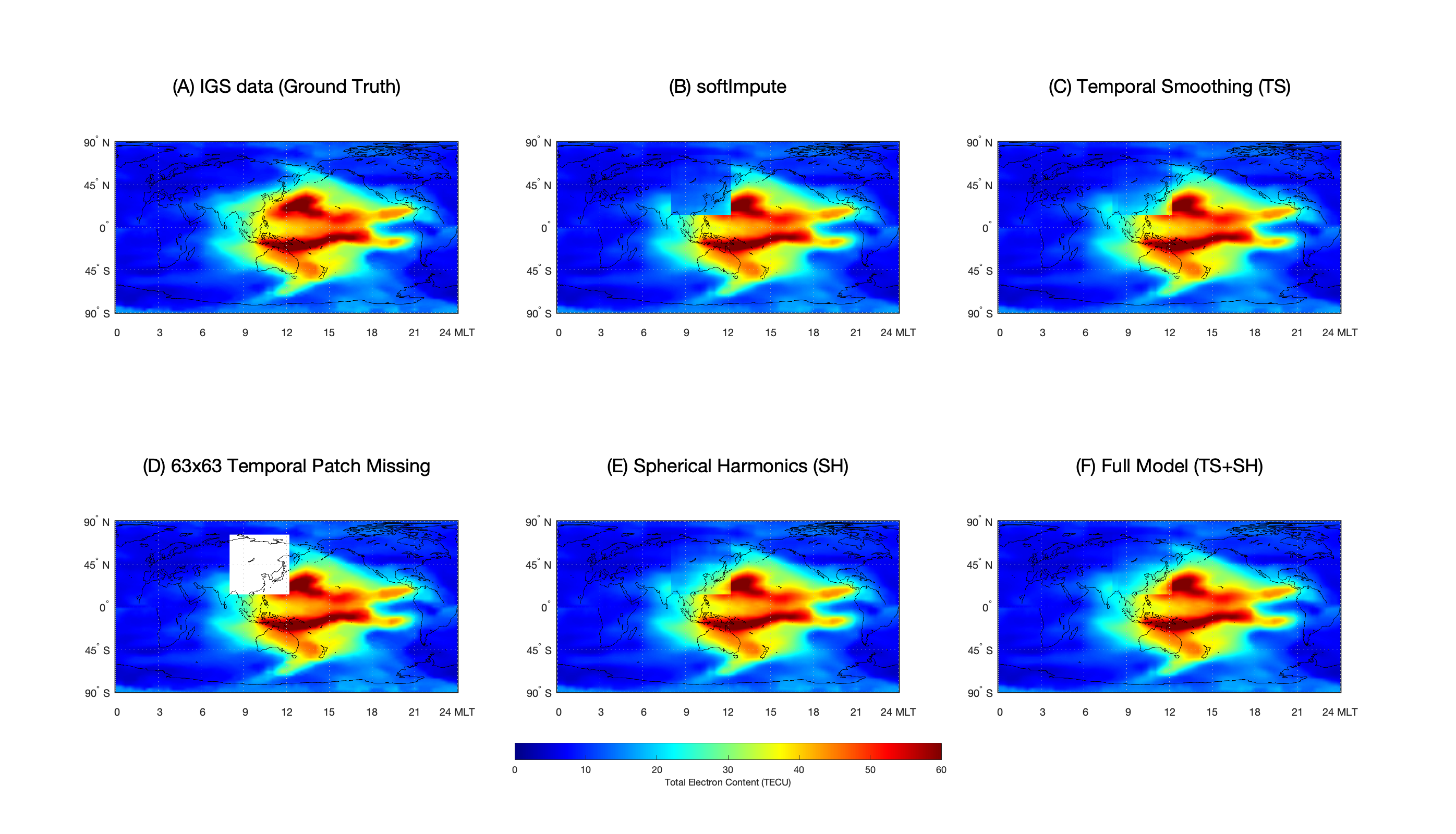}
    \caption{Example of imputing IGS data with temporal patch missingness. (A) IGS data at 2017-09-08 02:15:00 UT. (B) Imputed with softImpute ($\lambda_1=0.9$). (C) Imputed with temporal smoothing ($\lambda_1=0.9, \lambda_2=0.05$). (D) $63\times 63$ patch missingness. (E) Imputed with spherical harmonics auxliary data ($\lambda_1=0.9, \lambda_3=0.01$). (F) Imputed with full model ($\lambda_1=0.9, \lambda_2=0.05, \lambda_3=0.01)$.}
    \label{fig:temporal-patch-missing}
\end{figure}

We now give a concrete example as an illustration of the numerical results shown in Figure \ref{fig:patch-result}. In Figure \ref{fig:temporal-patch-missing}, we show the imputation made by 4 models when we have $63\times 63$ temporal patch missingness. It is easy to tell that softImpute barely imputes anything but background value in the patch. With temporal smoothing, the left and right borders of the patch are imputed. These are exactly the regions which are fully observed in its previous and next matrix. When using the spherical harmonics based auxiliary data, however, a more reasonable imputation is given in the patch. 

With the numerical analyses, we have illustrated how different variations of our model can outperform softImpute, the baseline method. Since the missing pattern in the TEC map of the madrigal database is mainly temporal patch missingness, it is expected that our extensions of the softImpute method can greatly help imputing this type of scientific images.

\subsection{Methodology Comparison}

Before concluding the numerical analysis section, we want to further compare our VISTA model against other competitive tensor completion methods beyond the SVD-based softImpute. We choose three methods: CP-WOPT \cite{acar2011scalable}, HaLRTC \cite{liu2012tensor} and TMac \cite{TMac}. The CP-WOPT is trying to find the best low-rank CP-decomposition of the imputation tensor so as to minimize the reconstruction error at the observed pixels. HaLRTC is using Alternating Direction Method of Multipliers (ADMM) to estimate the imputation tensor so as to minimize the weighted nuclear norm of the matricized data tensor. TMac is trying to find the best low-rank factorization the matricized data tensor and use the estimated factors to reconstruct the imputation.

For CP-WOPT, we set the $r=10$ (rank of the imputation tensor based on CP-decomposition). For HaLRTC, we set the $\alpha_{n} = \frac{1}{3}, \forall n$, so that each mode of the tensor gets equal weight in the objective function. For TMac, we set the targeting rank of each mode as $61$ (latitude), $64$ (magnetic local time), $96$ (temporal), respectively, based on the rank of the ground truth tensor after doing matricization along certain mode, and fit using a rank-increasing strategy. All hyper-parameter choices can be considered fair because they are chosen based on either the default or the recommended approach. Alongside these models, we also fit VISTA with $\lambda_1=0.5, \lambda_2=0.2, \lambda_3=0.04$ and softImpute with $\lambda=0.5$ based on the sequential tuning procedure described in section \ref{algorithm-section}. All simulation scenarios follow the previous setting and the test set RSE and error bar are shown in Figure \ref{fig:compare_tensor_completion}:

\begin{figure}[h]
    \centering
    \includegraphics[width=0.98\textwidth]{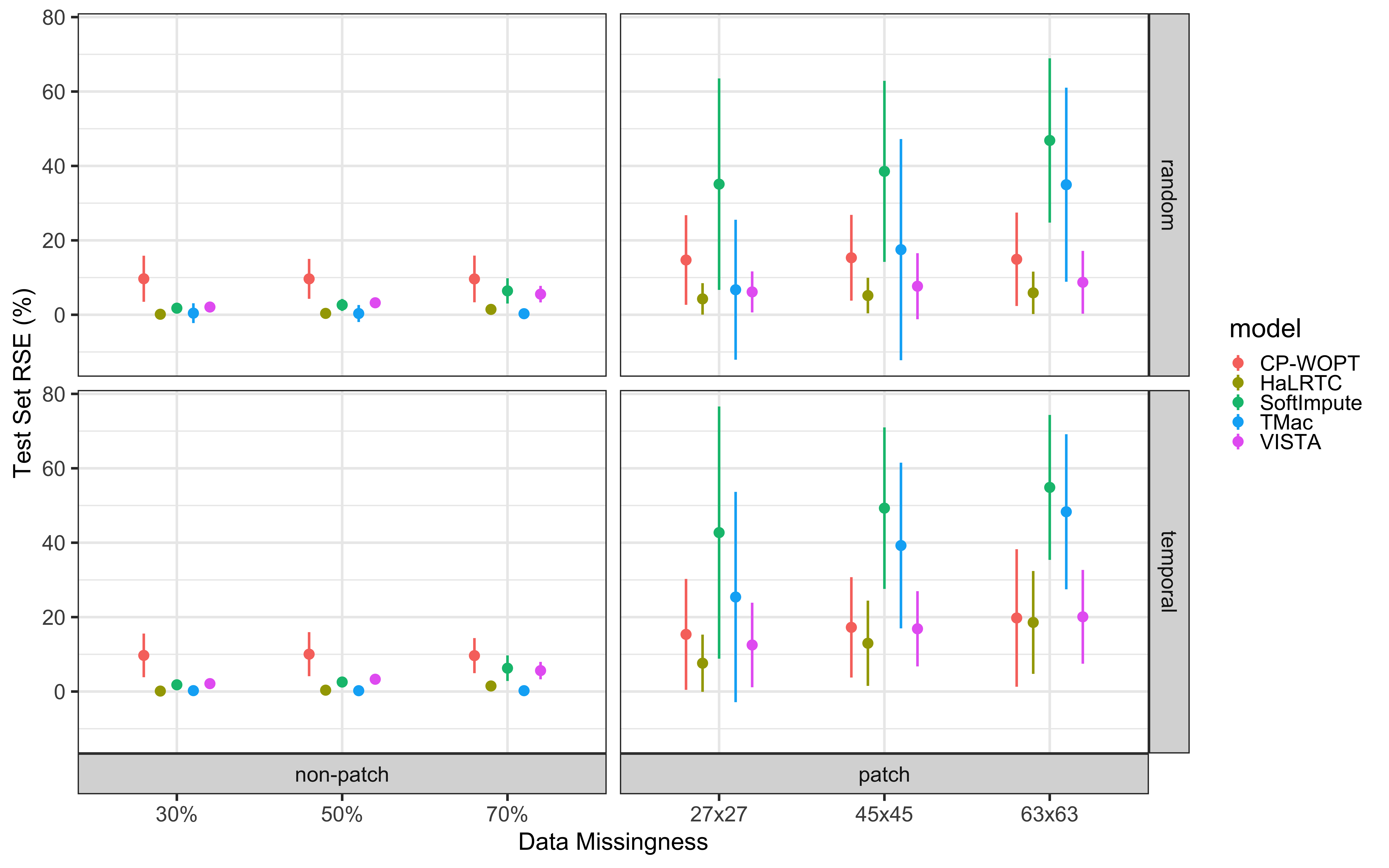}
    \caption{Method comparisons with CP-WOPT \cite{acar2011scalable}, HaLRTC \cite{liu2012tensor}, TMac \cite{TMac}, softImpute \cite{hastie2015matrix} and our VISTA. All four simulation scenarios are tested: random, temporal, patch and temporal patch. Three levels of data missingness are tested for each scenario. Choices of hyper-parameters are explained in texts. Error bar gives the $95\%$ confidence interval.}
    \label{fig:compare_tensor_completion}
\end{figure}

Figure \ref{fig:compare_tensor_completion} shows that our method is competitive against other selected methods in all four simulation scenarios and across all three levels of data missingness. The method that comes close to our VISTA model is the HaLRTC, and both have short error bars, indicating their consistency across different frames. In Figure \ref{fig:compare_tensor_completion_byframe}, we showed the test set RSE, by frame, for the simulation scenario of the temporal patch missingness:

\begin{figure}[h]
    \centering
    \includegraphics[width=0.85\textwidth]{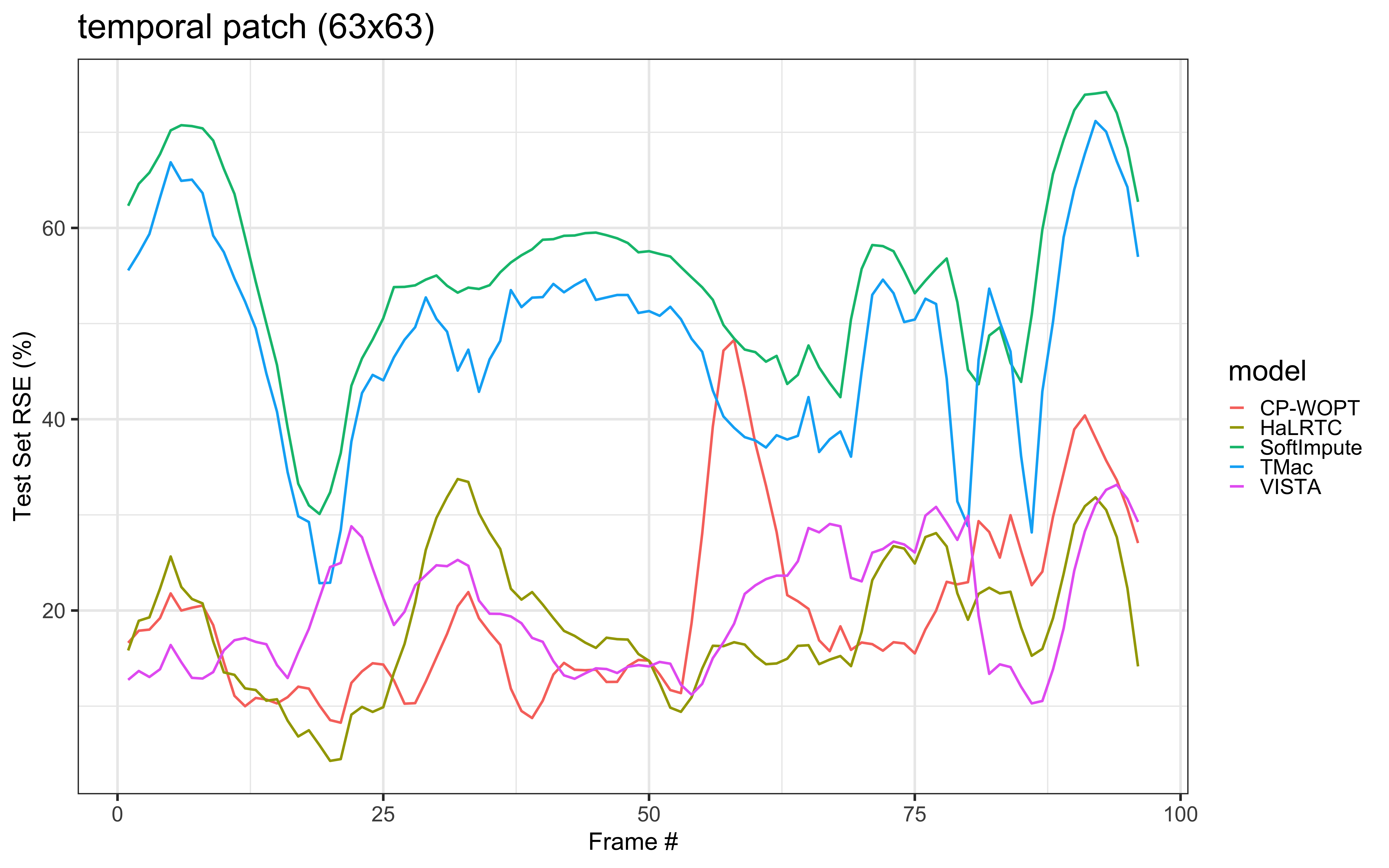}
    \caption{Method comparisons with CP-WOPT \cite{acar2011scalable}, HaLRTC \cite{liu2012tensor}, TMac \cite{TMac}, softImpute \cite{hastie2015matrix} and our VISTA for the temporal patch scenario with a $63\times 63$ box missing. The test set RSE is plotted against the frame number of the simulation data ($96$ frames in total).}
    \label{fig:compare_tensor_completion_byframe}
\end{figure}

We draw the conclusion from the Figure \ref{fig:compare_tensor_completion} and \ref{fig:compare_tensor_completion_byframe} that our VISTA model is comparable to the HaLRTC method and better than other methods when there are temporal patch missingness, which is the dominant type of missingness in the real TEC data. Also, when compared to HaLRTC, each method has some frames that does a better imputation. We recommend our approach for the ease of hyper-parameter tuning since based on our experience, VISTA is capable of providing reasonable imputations for a wide range of $\lambda_1,\lambda_2,\lambda_3$, and does not require information regarding the rank of the tensor at first place. VISTA also allows for other auxiliary data to play a role, which has extra possibilities of improving the imputation. It is an encouraging result to see that by combining Spherical Harmonics with softImpute, one can bring SVD-based method close or better to many competitive tensor completion method, given the bad performance of softImpute in the patch missing scenarios.

There are drawbacks for our method too, especially in terms of the computational efficiency. Take the temporal patch missingness ($63\times 63$) as an example, the HaLRTC takes $46.1$ seconds to terminate, the CP-WOPT takes $144.2$ seconds, the TMac takes $149.1$ seconds, while our VISTA takes $220$ seconds and there will be additional time consumed to generate the auxiliary data. All running times are recorded on a single-core (i9, 2.3 GHz), $16$-GB memory (2400MHz, DDR4) CPU. Other simulation scenarios take similar time to finish for each of the algorithm. We admit that our model is not as efficient and scalable as the others, but we want to highlight that our VISTA model can generate similar imputation result for a wide range of tuning parameters, leading to potentially shorter running time if one counts the cross-validation step too.

In the next section, we present our empirical results on imputing the TEC map of madrigal database and address how our extensions can solve the problems of using softImpute for TEC map imputation, as discussed in Section~\ref{subsec:litreview}.

\section{TEC Map Reconstruction Results}

We apply our model to reconstruct the sparse TEC maps in this section. For each day of the TEC data, our data has the form $X = \{X_{ijk}\} \in \mathbb{R}^{m \times n\times T}$, where $m = 181$ corresponds to the latitude, $n = 361$ corresponds to the longitude and $T = 288$ corresponds to the number of time points of the day. We use TEC median filtered data in this section. Specifically, we choose two days data for reconstruction demonstration: 2017-09-08 (a storm day) and 2017-09-03 (a non-storm day). During a storm day, the TEC map has more spatial structures and is more variable than that during a non-storm day.

Each whole day of the data contain 288 matrices, and in each matrix, we randomly drop $20\%$ of the observed pixels and use them as the testing set. The remaining 80\% of the data is used as the training set. We fit all 288 matrices together in one algorithm run and validate model performances on the testing set. In this section, data follows the same pipeline as shown in Figure \ref{fig:pipeline}.

Hyper-parameters are selected based on the relative square error (RSE) score. First we set $\lambda_2 = 0, \lambda_3 = 0$ and perform grid search on $\lambda_1$ to get the best $\lambda_1$ value. Again, this choice of $\lambda_1$ optimizes the performance of the original softImpute algorithm. Then, with our best $\lambda_1$ value, we perform grid search on temporal smoothing term and SH term separately for the best $\lambda_2$ and $\lambda_3$ values. Although this choice of the set of tuning parameters could be sub-optimal, it is computationally much more efficient and well serves the purpose of comparisons of various algorithms. 

Table \ref{tab:emp_res} shows the results of our model on a storm day and a non-storm day. For the storm day, the best temporal smoothing hyper-parameter is $\lambda_2 = 0.2$ and the best spherical harmonic hyper-parameter is $\lambda_3 = 0.021$. The lowest RSE is achieved when we use the full model with $(\lambda_1, \lambda_2, \lambda_3) = (0.9, 0.2, 0.021)$. RSE reduces from 10.895\% achieved by softImpute to 9.357\% achieved by the full model. For the non-storm day, the best temporal smoothing hyper-parameter is $\lambda_2 = 0.31$ and the best spherical harmonic hyper-parameter is $\lambda_3 = 0.03$. When we use the full model with $(\lambda_1, \lambda_2, \lambda_3) = (0.9, 0.31, 0.03)$, we achieve lowest RSE. RSE reduces from 10.424\% achieved by softImpute to 8.592\% achieved by the full model. Moreover, we also examined the mean-squared error (MSE) on the testing set. It turns out that the full model MSE also tends to be lower than other models, and the softImpute model gives the highest MSE value among all the other imputation models we compare. The MSE of the non-storm day is overall lower than that of the storm day. This might be attributed to the relatively lower TEC values on a non-storm day. This also indicates that it is easier to impute non-storm day data, which varies less and has lower magnitudes. Based on the last two columns of Table~\ref{tab:emp_res}, in over 97.5\% of the time points, the TS, SH and full model perform better than the softImpute result. And in over 81\% of the time points, the full model outperforms the other models, indicating that both temporal smoothing and auxiliary data from spherical harmonics can help on improving the imputation results of TEC maps.

\begin{table}[h]
    \centering
    \begin{tabular}{c|c|c|c|c}
    \hline
    \multicolumn{5}{c}{Storm Day} \\
    \hline
    Model & test RSE & test MSE & \makecell[c]{\# matrices better \\ than softImpute} & \makecell[c]{
    \# matrices worse \\ than Full model}\\
    \hline
      softImpute ($\lambda_1=0.9$) & $10.895\%$ & 2.675 & /  & 285 (98.96\%)\\
      TS ($\lambda_1=0.9, \lambda_2=0.2$) & $9.643\%$ & 2.106 & 284 (98.62\%) & 267 (92.71\%) \\
      SH ($\lambda_1=0.9, \lambda_3=0.021$) & $9.936\%$ & 2.227 & 287 (99.65\%) & 274 (95.14\%) \\
      Full ($\lambda_1=0.9, \lambda_2=0.2, \lambda_3=0.021$)  & $9.357\%$ & 1.983 & 285 (98.96\%) & /\\
      Directly use Spherical Harmonics& $17.354\%$ & 6.720 & 0 (0\%) & 288 (100\%)\\\hline
      \multicolumn{5}{c}{Non-Storm Day} \\
    \hline
    Model & test RSE & test MSE & \makecell[c]{\# matrices better \\ than softImpute} &  \makecell[c]{\# matrices worse \\ than Full model}\\
    \hline
      softImpute ($\lambda_1=0.9$) & $10.424\%$ & 1.324 & /  & 283 (98.26\%)  \\
      TS ($\lambda_1=0.9, \lambda_2=0.31$) & $8.880\%$ & 0.958 &  281 (97.57\%) & 235 (81.60\%) \\
      SH ($\lambda_1=0.9, \lambda_3=0.03$) & $9.231\%$ & 1.032 & 287 (99.65\%) & 278 (96.53\%) \\
      Full ($\lambda_1=0.9, \lambda_2=0.31, \lambda_3=0.03$) & $8.592\%$ & 0.895 & 283 (98.26\%) & /\\
      Directly use Spherical Harmonics& $15.732\%$ & 2.893 & 0 (0\%) & 288 (100\%)\\\hline
    \end{tabular}
    \caption{Empirical study results from the madrigal database.}
    \label{tab:emp_res}
\end{table}

\begin{figure}[h]
    \centering
    \includegraphics[width=0.98\textwidth]{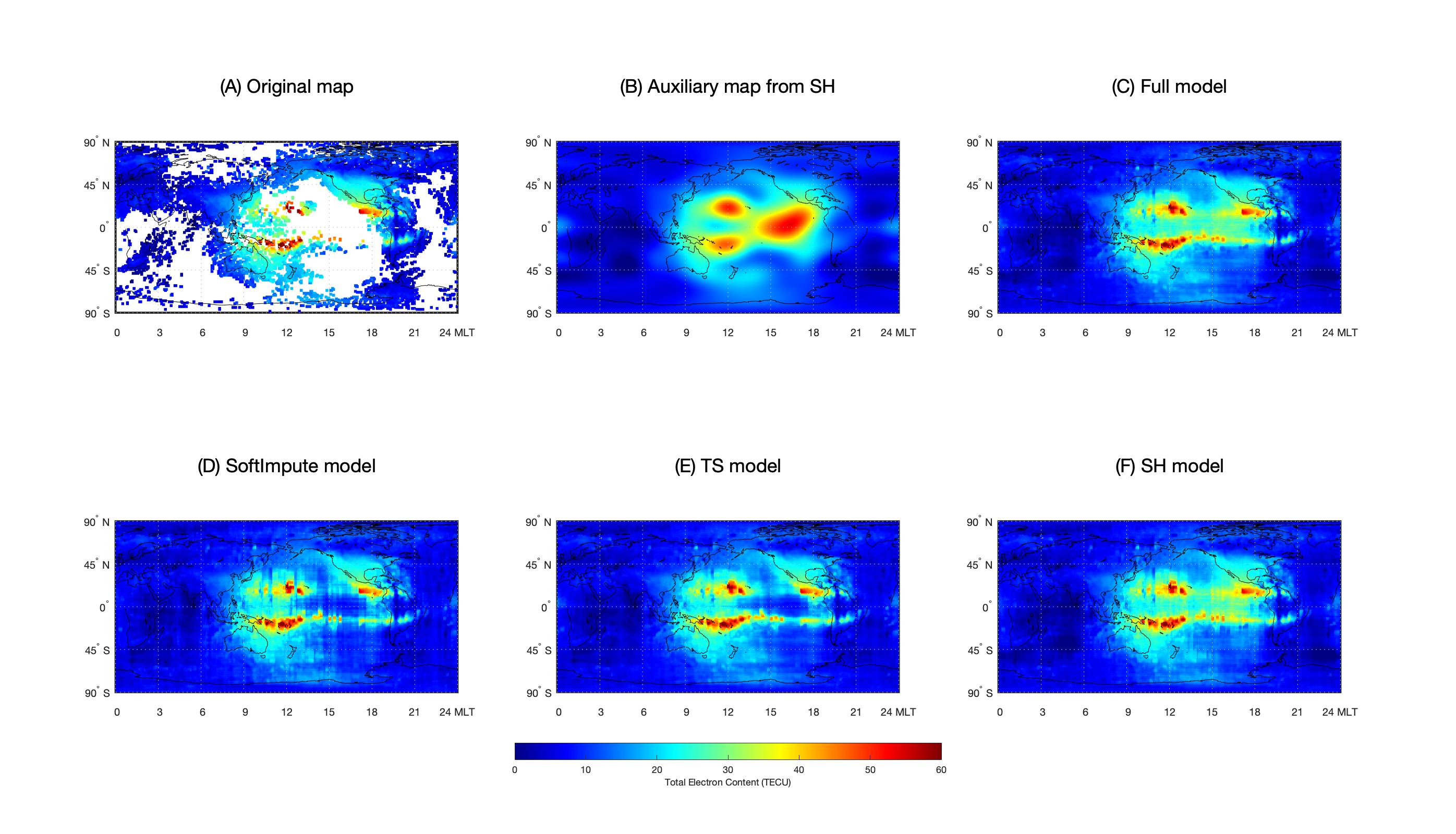}
    \caption{2017-09-08/00:02:30 UT geomagnetic local time maps. (A) Original median filter map. (B) Directly use SH fitted map. (C) Full model imputed map with $(\lambda_1, \lambda_2, \lambda_3) = (0.9, 0.2, 0.021)$. (D) SoftImpute fitted map with $\lambda_1 = 0.9$. (E) Temporal smoothing imputed map with $(\lambda_1, \lambda_2) = (0.9, 0.2)$. (F) SH imputed map with $(\lambda_1, \lambda_3) = (0.9, 0.021)$.}
    \label{fig:0908}
\end{figure}

\begin{figure}[h]
    \centering
    \includegraphics[width=0.98\textwidth]{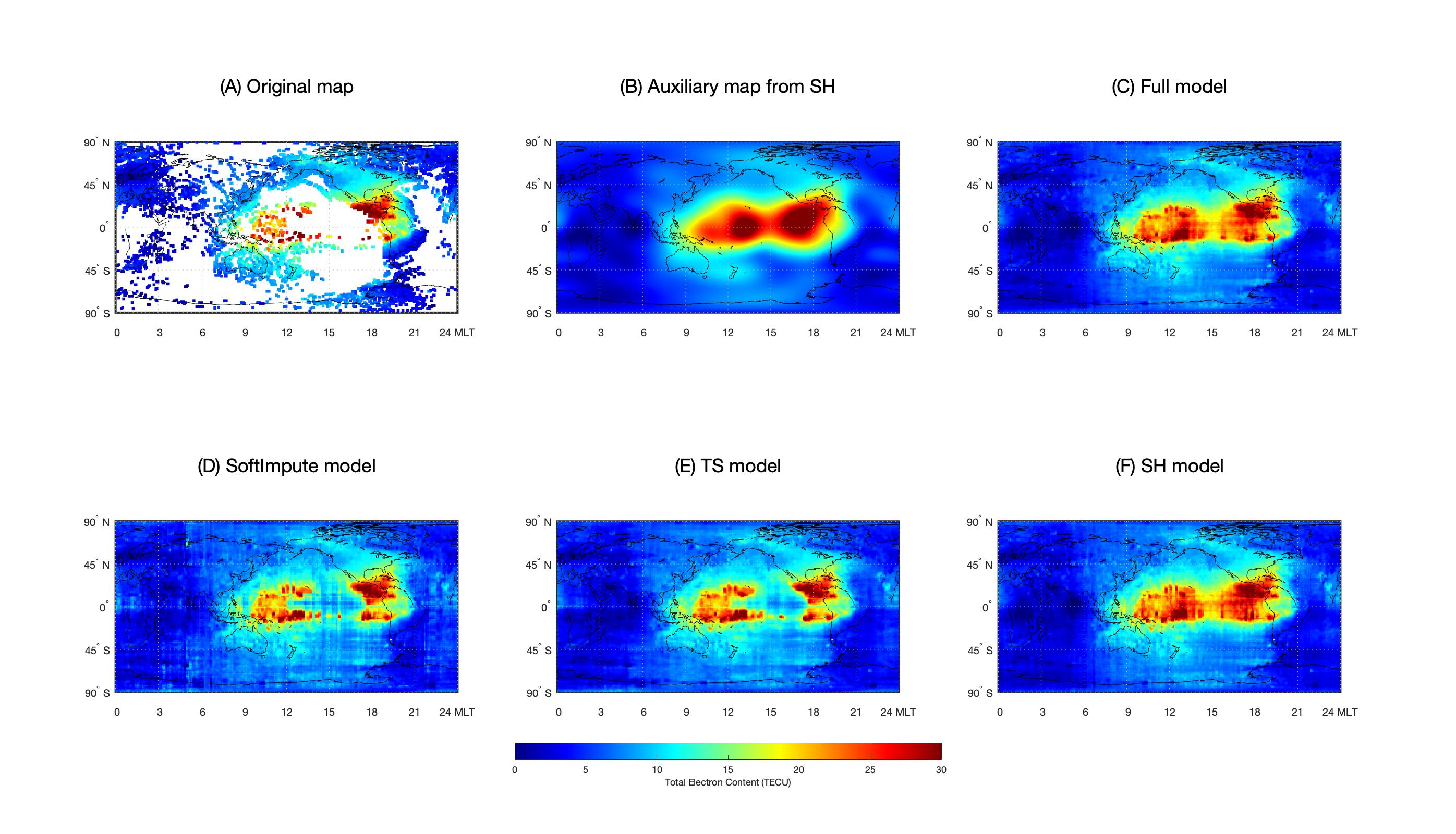}
    \caption{2017-09-03/00:02:30 UT geomagnetic local time maps. (A) Original median filter map. (B) Directly use SH fitted map. (C) Full model imputed map with $(\lambda_1, \lambda_2, \lambda_3) = (0.9, 0.31, 0.03)$. (D) SoftImpute fitted map with $\lambda_1 = 0.9$. (E) Temporal smoothing imputed map with $(\lambda_1, \lambda_2) = (0.9, 0.31)$. (F) SH imputed map with $(\lambda_1, \lambda_3) = (0.9, 0.03)$.}
    \label{fig:0903}
\end{figure}

Figure \ref{fig:0908} and \ref{fig:0903} show the original TEC median filtered map, the map fitted with spherical harmonics (auxiliary data) and four imputed maps of one time point of the storm day and the non-storm day data, respectively. The original softImpute method with only $\lambda_1$ value imposes a low-rank structure on the imputed matrix, leading to unreasonable gaps when being applied to scientific images. Our method with all three hyper-parameters reserves the features of the original plot, mitigates such gaps to allow for more temporal consistency and gives better spatial smoothness. {For a non-storm day, the average rank of the imputed map with softImpute is $72$, with temporal-smoothing is $98.5$, and with the full model is $99.8$. For a storm day, the average rank of the imputed map with softImpute is $85$, with temporal smoothing is $104$, and with the full model is $105.4$.} Note that Figures \ref{fig:0908} and \ref{fig:0903} have different scales due to the relatively lower TEC value on a non-storm day. 

Comparing Figure \ref{fig:0908} (D)(E) or (F)(C), we can see that by adding temporal smoothing penalty, undesirable low-rank structure is smoothed out. Comparing \ref{fig:0908} (D)(F) or (E)(C), the spherical harmonics helps filling in missing patches caused by a lack of observations which appear in blue patches in (D) and (E) near the equator, i.e. 18 magnetic local time (MLT). The same patterns can be found in Figure \ref{fig:0903}.

Overall, one can see that the temporal smoothing helps eliminate the low-rank structure; and the auxiliary data based on spherical harmonics, helps fill in large missing patches. These are exactly the two drawbacks of using the original trace-norm based matrix completion method for imputing TEC maps.

Just like the numerical analysis section, we also want to report the computational time for real TEC map reconstruction. For a single-core (i9, 2.3 GHz), $16$-GB memory (2400MHz, DDR4) CPU, and for the storm day data($181\times 361\times 288$), the HaLRTC takes $112.3$s, the CP-WOPT takes $1522.5$s, the TMac takes $409.8$s, while our VISTA model takes $1134.5$s to terminate. The computational efficiency and scalability of our method is not ideal, but our current version is still a workable solution to the TEC map and similar video imputation problem.

Our work goes beyond methodology research and can contribute to generating high-quality TEC maps for the scientific field too. Meso-scale ionospheric structures, such as the channel-like TEC depletions in the north-south direction at ~20 LT in Figure \ref{fig:madrigal} and Figure \ref{fig:0908} (i.e., equatorial plasma bubble [e.g. \cite{aa2018midlatitude,aa2019,abdu2019day} and references therein]) are associated with the largest navigation and communication satellite signal scintillations and thus of essential practical significance. However, this type of meso-scale structures can’t be captured in the standard SH harmonics fitted maps, such as the IGS TEC maps. Therefore, the capability of preserving these meso-scale structures in the completed TEC map is a must for us to improve our specification and forecast of ionospheric space weather impact and will enable numerous studies in the domain field, such as the development and evolution of these meso-scale structures, their interaction with large-scale TEC structures (i.e., the storm-enhanced density \cite{foster2005multiradar,zou2013electrodynamics,Zou2014,zou2016modeling} ), as well as the forecast of TEC using machine learning techniques based on these new maps. Our VISTA model gives an option to take a solid first step on these research tracks.

\section{Conclusion}

In this paper, we proposed the Video Imputation with SoftImpute, Temporal smoothing and Auxiliary data (VISTA) method, which gives two extensions of the softImpute algorithm for matrix completion in temporal TEC map reconstruction. The incorporation of auxiliary data and temporal smoothness via penalty terms in the loss function enables us to combine external information and achieve temporal information sharing. The proposed algorithm is implemented in R and made available on Github. An R package is currently under preparation. We prove theoretical properties of the algorithm such as the convergence rate as in the original softImpute paper. In our numerical simulations that mimic the real world scenario, the proposed method works out very well and vastly improves the performance of the SVD-based method. The real data analysis further demonstrate satisfactory performance.

There are a few extensions of our method that may trigger further research interests. For instance, when one creates the majorization bound, more generally, one can:

\begin{equation}\label{MM_extension}
    \|P_{\Omega_t}(X_t - A_{t}(B_{t}^{(k)})^T)\|_F^2 \le \|P_{\Omega_t}(X_t) + a P_{\Omega_t^{\perp}} (A_{t}^{(k)}(B_{t}^{(k)})^T) - A_{t}(B_{t}^{(k)})^T\|_F^2.
\end{equation}

We found that by letting $a<1$, the algorithm still converges and the smaller $a$ is, the quicker the algorithm converges. However, given the same tuning parameter, $a=1$ still has better empirical performance on real TEC data. But this still suggests more room for new research on the majorization bound to further improve these type of method.

The output of the imputed TEC data suggests that the rank of the imputed map, for a single frame, is around $80\sim 120$, while the maximum possible rank should be $181$. This suggests that the true underlying map can have a low-rank nature, which can be modelled more directly via non-convex matrix completion method. We adopt the softImpute-ALS for its convexity and its flexibility to add temporal smoothness and auxiliary data penalty. More researches on how to combine these penalties with a non-convex matrix completion framework can be beneficial, especially when one has prior knowledge about the low-rank nature of the data.

Another extension of our framework is to localize the temporal smoothness and auxiliary data penalty to a specific spatial-temporal patch. Our current setup has both penalty for all frames and all pixels, which can be sub-optimal. We found that the optimal tuning parameters for storm and non-storm day differ, which suggests that different time can have different temporal/spatial smoothness, indicating the significance of more localized spatial-temporal smoothness penalty. More researches on this thread to further extend our framework and similar tensor-completion method can be highly beneficial.

The last extension is about the selection and generation of auxiliary data. In this paper, we adopt the spherical harmonics as the auxiliary data and fit it separately from our main model. One can also select the outputs of other imputation methods (e.g. HaLRTC, CP-WOPT, etc.) as the auxiliary data alongside the spherical harmonics. The selection should be based on the real data. Spherical harmonics is a good choice for TEC data given its nature, but is not necessarily good for other types of data. Also, one can fit the auxiliary data jointly with the imputation model. We do not do it here since more tuning parameters will get involved if one fits the model jointly, but it is definitely a possibility to generate the auxiliary data together with the imputation map.

The newly proposed algorithms is targeted but not restricted to the temporal TEC map reconstruction. We hope that the proposed general algorithm can serve a larger scientific community and stimulate more interest from statisticians for more thorough theoretical investigations.

\section*{Acknowledgement}
This work (SZ, YC, and JR) is supported by NASA DRIVE Center at the University of Michigan under grant NASA 80NSSC20K0600. SZ acknowledges NASA award 80NSSC20K1313 and 80NSSC20K0190. YC acknowledges support from NSF DMS Award 1811083 and 2113397. SZ and YC also acknowledges NSF PHY Award 2027555.

\section*{Supplement}

\section{Proof of Theoretical Results in Section 2.3}
\subsection{Proof of Theorem 2.1}\label{proof-descent-theorem}
\begin{proof}
The objective function $F(A_{1:T}^{(k)},B_{1:T}^{(k)})$ has the property:
\begin{align*}
    F(A_{1:T}^{(k)},B_{1:T}^{(k)}) & = \tilde{Q}(A_1^{(k)}|A_{1:T}^{(k)},B_{1:T}^{(k)})\\
    & \geq \inf_{A_1} \tilde{Q}(A_1|A_{1:T}^{(k)},B_{1:T}^{(k)}) \\
    & = \tilde{Q}(A_1^{(k+1)}|A_{1:T}^{(k)},B_{1:T}^{(k)}) \label{T1-step1}\numberthis{}\\
    & \geq F(A_1^{(k+1)},A_{2:T}^{(k)},B_{1:T}^{(k)}), \label{T1-step2}\numberthis{}
\end{align*}
where the definition of $\tilde{Q}$ is in (9) of the paper. Equation \eqref{T1-step1} holds because we update $A_1$ to be $A_1^{(k+1)}$ using ridge regression: $A_1^{(k+1)} = \argmin  \tilde{Q}(A_{1}|A_{1:T}^{(k)},B_{1:T}^{(k)})$. Inequality \eqref{T1-step2} holds because $\tilde{Q}(A_1^{(k+1)}|A_{1:T}^{(k)},B_{1:T}^{(k)})$ is the upper bound of $F(A_1^{(k+1)},A_{2:T}^{(k)},B_{1:T}^{(k)})$, as we majorize the first term of the objective function using inequality (6) of the paper. 

The property above indicates that after one single update of matrix $A_1$, the values of the objective function is non-increasing. Applying a similar argument for all other matrices $A_2, A_3, \dots, A_T, B_1, B_2, \dots, B_T$ leads to a chain of inequalities:
\begin{align*}
    F(A_{1:T}^{(k)},B_{1:T}^{(k)}) \geq F(A_1^{(k+1)},A_{2:T}^{(k)},B_{1:T}^{(k)}) \geq F(A_{1:2}^{(k+1)},A_{3:T}^{(k)},B_{1:T}^{(k)}) \geq \dots \geq F(A_{1:T}^{(k+1)},B_{1:T}^{(k)}) \\ \geq F(A_{1:T}^{(k+1)},B_1^{(k)},B_{2:T}^{(k)})\geq F(A_{1:T}^{(k+1)},B_{1:2}^{(k)},B_{3:T}^{(k)}) \geq \dots F(A_{1:T}^{(k+1)},B_{1:T}^{(k+1)}),
\end{align*}
which proves that the each update of $A_t$ or $B_t$ goes towards a descent direction.
\end{proof}

\subsection{Proof of Theorem 2.2}\label{proof-descent-LB}
\begin{proof}
Note that in appendix \ref{proof-descent-theorem}, we proved inequality \eqref{T1-step2}. More generally, for any arbitrary $t$, we have the following:
\begin{align}\label{T2-step1}
    \Delta^{A}_{k,t} \triangleq &F(A_{1:t-1}^{(k+1)},A_{t:T}^{(k)},B_{1:T}^{(k)}) -  F(A_{1:t}^{(k+1)},A_{t+1:T}^{(k)},B_{1:T}^{(k)}) \\
    \geq &\tilde{Q}(A_{t}^{(k)}|A_{1:t-1}^{(k+1)},A_{t:T}^{(k)},B_{1:T}^{(k)}) - \tilde{Q}(A_{t}^{(k+1)}|A_{1:t-1}^{(k+1)},A_{t:T}^{(k)},B_{1:T}^{(k)}).\nonumber
\end{align}
The right hand side of \eqref{T2-step1} is the difference of $\tilde{Q}(A_{t}|A_{1:t-1}^{(k+1)},A_{t:T}^{(k)},B_{1:T}^{(k)})$ evaluated at $A_t^{(k)}$ and $A_t^{(k+1)}$. Recall that:
\begin{align*}
    \tilde{Q}(A_{t}|A_{1:t-1}^{(k+1)},A_{t:T}^{(k)},B_{1:T}^{(k)}) &\triangleq \frac12\|X^{(k)}_t - A_{t}(B_{t}^{(k)})^T\|_F^2 + \frac{\lambda_1}{2}\|A_t\|_F^2 + \frac{\lambda_3}{2} \|Y_t - A_{t}(B_{t}^{(k)})^T\|_F^2\\
    &\quad + \frac{\lambda_2}{2}\mathbf{I}_{\{t> 1\}}\|A_{t}(B_{t}^{(k)})^T - A_{t-1}^{(k+1)}(B_{t-1}^{(k)})^T\|_F^2 \\
    &\quad + \frac{\lambda_2}{2}\mathbf{I}_{\{t< T\}}\|A_{t+1}^{(k)}(B_{t+1}^{(k)})^T - A_{t}(B_{t}^{(k)})^T\|_F^2.
\end{align*}
Note that this is a quadratic function of $A_t$ thus higher order ($\geq 3$) derivatives are all zero. We can do a Taylor expansion for $\tilde{Q}(A_{t}^{(k)}|A_{1:t-1}^{(k+1)},A_{t:T}^{(k)},B_{1:T}^{(k)})$ at $A_t^{(k+1)}$:
\begin{align*}\label{T2-step2}
   \tilde{Q}(A_{t}^{(k)}|A_{1:t-1}^{(k+1)},A_{t:T}^{(k)},B_{1:T}^{(k)}) & =  \tilde{Q}(A_{t}^{(k+1)}|A_{1:t-1}^{(k+1)},A_{t:T}^{(k)},B_{1:T}^{(k)}) \\
   & + (\nabla \tilde{Q}) (A_{t}^{(k)} - A_{t}^{(k+1)})\\
   &+ \frac12 (A_{t}^{(k)} - A_{t}^{(k+1)})^T H (A_{t}^{(k)} - A_{t}^{(k+1)}), \numberthis{}
\end{align*}
where $H = (1+\lambda_2(1+\mathbf{I}_{\{2\leq t\leq T-1\}})+\lambda_3)(B_{t}^{(k)})^TB_{t}^{(k)} + \lambda_1 I$. We have $\nabla\tilde{Q} = 0$ since $A_t^{(k+1)}$ is the minimizer of $ \tilde{Q}(A_{t}|A_{1:t-1}^{(k+1)},A_{t:T}^{(k)},B_{1:T}^{(k)})$. Combining \eqref{T2-step1} and \eqref{T2-step2}, one can see that:
\begin{align*}\label{T2-step3}
    \Delta^{A}_{k,t}  & \geq \tilde{Q}(A_{t}^{(k)}|A_{1:t-1}^{(k+1)},A_{t:T}^{(k)},B_{1:T}^{(k)}) - \tilde{Q}(A_{t}^{(k+1)}|A_{1:t-1}^{(k+1)},A_{t:T}^{(k)},B_{1:T}^{(k)})\\
    & = \frac12 (A_{t}^{(k)} - A_{t}^{(k+1)})^T H (A_{t}^{(k)} - A_{t}^{(k+1)}) \\
    & = \frac{1+\lambda_2(1+\mathbf{I}_{\{2\leq t\leq T-1\}})+\lambda_3}{2} \|(A_{t}^{(k)} - A_{t}^{(k+1)})(B_{t}^{(k)})^T\|^2 \\
    & + \frac{\lambda_1}{2} \|A_{t}^{(k)} - A_{t}^{(k+1)}\|^2. \numberthis{}
\end{align*}
Similarly for any updates of $B_t$, we have:
\begin{align*}\label{T2-step4}
    \Delta_{k,t}^{B} &\triangleq  F(A_{1:T}^{(k+1)},B_{1:t-1}^{(k+1)},B_{t:T}^{(k)}) -  F(A_{1:T}^{(k+1)},B_{1:t}^{(k+1)},B_{t+1:T}^{(k)})\\
    & \geq \tilde{Q}(B_{t}^{(k)}|A_{1:T}^{(k+1)},B_{1:t-1}^{(k+1)},B_{t:T}^{(k)}) - \tilde{Q}(B_{t}^{(k+1)}|A_{1:T}^{(k+1)},B_{1:t-1}^{(k+1)},B_{t:T}^{(k)})\\
    & = \frac{1+\lambda_2(1+\mathbf{I}_{\{2\leq t\leq T-1\}})+\lambda_3}{2} \|A_{t}^{(k+1)}(B_{t}^{(k)} - B_{t}^{(k+1)})^T\|^2 \\
    & + \frac{\lambda_1}{2} \|B_{t}^{(k)} - B_{t}^{(k+1)}\|^2. \numberthis{}
\end{align*}

Since \eqref{T2-step3} and \eqref{T2-step4} hold for all $A_t$ and $B_t$, we can sum the $\Delta_{k,t}^{A}, \Delta_{k,t}^{B}$ across all $t$. Note that $\sum_t (\Delta_{k,t}^{A} + \Delta_{k,t}^{B}) = \Delta_k$. The right-hand side is the lower bound for $\Delta_{k}$ that we want.
\end{proof}

\subsection{Proof of Theorem 2.3}\label{proof-conv-rate-theorem}
\begin{proof}
The first result can be easily proved by noting that
\begin{equation}\label{T3-step1}
    F(A_{1:T}^{(1)},B_{1:T}^{(1)}) - f^{\infty} \geq F(A_{1:T}^{(1)},B_{1:T}^{(1)}) - F(A_{1:T}^{(K)},B_{1:T}^{(K)}) = \sum_{k=1}^K \Delta_k \geq K(\min_{1\leq k\leq K}\Delta_k).
\end{equation}
Given the assumption that $l^{L}\mathrm{I}\leq(A_t^{(k)})^{T}A_t^{(k)}\leq l^{U}\mathrm{I}$, $l^{L}\mathrm{I}\leq(B_t^{(k)})^{T}B_t^{(k)}\leq l^{U}\mathrm{I}$ for all $t,k$. Equations (18) and (19) of the paper can be proved with the following inequalities:
\begin{align}
    l^{L}\|A_{t}^{(k)} - A_{t}^{(k+1)}\|^2 \leq \|(A_{t}^{(k)} - A_{t}^{(k+1)})(B_{t}^{(k)})^T\|^2 \leq l^{U}\|A_{t}^{(k)} - A_{t}^{(k+1)}\|^2; \label{T3-step2}\\
    l^{L}\|B_{t}^{(k)} - B_{t}^{(k+1)}\|^2 \leq \|A_t^{(k+1)}(B_{t}^{(k)} - B_{t}^{(k+1)})^T\|^2 \leq l^{U}\|B_{t}^{(k)} - B_{t}^{(k+1)}\|^2.\label{T3-step3}
\end{align}
Given the lower bound in theorem 2.2 and the inequality in \eqref{T3-step1}, we have:
\begin{align*}
    & \frac{F(A^{(1)}_{1:T},B^{(1)}_{1:T}) - f^{\infty}}{K}  \geq \min_{1\leq k\leq K}\Delta_k \\
    & \geq \min_{1\leq k\leq K} \bigg\{\frac{\lambda_1}{2}\sum_{t=1}^T \left(\|A_t^{(k)}-A_t^{(k+1)}\|^2 + \|B_t^{(k)}-B_t^{(k+1)}\|^2 \right)\\
    & + \frac{1}{2} \sum_{t=1}^T (1+\lambda_2+\lambda_3) \left(\|(A_t^{(k)}-A_t^{(k+1)})(B_t^{(k)})^T\|^2 + \|A_t^{(k+1)}(B_t^{(k)}-B_t^{(k+1)})^T\|^2\right)\bigg\} \\
    & \geq \min_{1\leq k\leq K} \left\{\frac{l^{L}(1+\lambda_2+\lambda_3)+\lambda_1}{2} \sum_{t=1}^T \left(\|A_t^{(k)}-A_t^{(k+1)}\|^2 + \|B_t^{(k)}-B_t^{(k+1)}\|^2 \right)\right\}.
\end{align*}
The last step uses the left inequality in \eqref{T3-step2} and \eqref{T3-step3}. This proves (18). Using the right-hand side inequality in \eqref{T3-step2} and \eqref{T3-step3} yields (19).
\end{proof}

\bibliographystyle{plain}
\bibliography{main.bib}
\end{document}